\documentclass[]{aa} 
\usepackage{natbib} 
\usepackage{graphicx}

\def\onesig{1-$\sigma$~level~}
\def\onesign{1-$\sigma$~level}
\def\rms{r{.}m{.}s{.}~}
\def\resp{resp{.}~}
\def\respn{resp{.}}

\def\eq{Eq{.}~}                 

\def\fg{Fig{.}~}
\def\fgs{Figs{.}~}
\def\sct{Sect{.}~}

\def\scts{Sects{.}~}
\def\sctsn{Sects{.}}
\def\sctsnp{Sects}
\def\col{Col{.}~}

\def\cols{Cols{.}~}

\def\phiunit{~$h^3$~Mpc$^{-3}$~mag$^{-1}$~}
\def\phiunitn{~$h^3$~Mpc$^{-3}$~mag$^{-1}$}

\begin{document}

\title{Critical analysis of the luminosity functions per galaxy type
measured from redshift surveys}

\author{Val\'erie de Lapparent}


\institute{Inst{.} d'Astrophysique de Paris, CNRS, Univ. Pierre et
Marie Curie, 98 bis Boulevard Arago, 75014 Paris, France\\
\email{lapparen@iap.fr} }

\date{Received 27 March 2003 /  Accepted 12 June 2003}

\abstract{ I perform a quantitative comparison of the shape of the
optical luminosity functions as a function of galaxy class and filter,
which have been obtained from redshift surveys with an effective depth
ranging from $z\simeq0.01$ to $z\simeq0.6$.  This analysis is based on
the $M^*$ and $\alpha$ Schechter parameters which are systematically
measured for all galaxy redshift surveys. I provide complete tables of
all the existing measurements, which I have converted into the
$UBVR_\mathrm{c}I_\mathrm{c}$ Johnson-Cousins system wherever
necessary.

\hspace{0.5cm} By using as reference the intrinsic luminosity
functions per morphological type, I establish that the variations in
the luminosity functions from survey to survey and among the galaxy
classes are closely related to the criteria for galaxy classification
used in the surveys, as these determine the amount of mixing of the
known morphological types within a given class. When using a spectral
classification, the effect can be acute in the case of inaccurate
spectrophotometric calibrations: the luminosity functions are then
biased by type contamination and display a smooth variation from type
to type which might be poorly related to the intrinsic luminosity
functions per morphological type. In the case of surveys using
multi-fiber spectroscopy, galaxy classification based on rest-frame
colors might provide better estimates of the intrinsic luminosity
functions.

\hspace{0.5cm} It is noticeable that all the existing redshift surveys
fail to measure the Gaussian luminosity function for Spiral galaxies,
presumably due to contamination by dwarf galaxies. Most existing
redshift surveys based on visual morphological classification also
appear to have their Elliptical/Lenticular luminosity functions
contaminated by dwarf galaxies. In contrast, the analyses using a
reliable spectral classification based on multi-slit spectroscopy or
medium-filter spectrophotometry, and combined with accurate CCD
photometry succeed in measuring the Gaussian luminosity function for
E/S0 galaxies.  The present analysis therefore calls for a more
coherent approach in separating the relevant giant and dwarf galaxy
types, a necessary step towards measuring reliable intrinsic
luminosity functions.

 }

\authorrunning{de Lapparent}
\titlerunning{Critical analysis of luminosity functions from redshift surveys}
\maketitle

\section{Introduction \label{intro}} 

Among the fundamental characteristics of galaxies is their luminosity
function (LF hereafter). In the current models of galaxy formation
based on gravitational clustering, the LF provides constraints on the
formation of galaxies within the dark matter halos
\citep{cole00,baugh02}, thus allowing one to adjust the parameters for
star formation, feedback processes, and mergers within the
halos. Based on assumptions about the formation of bulge-dominated and
disk-dominated galaxies, the various galaxy types can be traced
separately in the models, which enables one to perform direct
comparison with the observations \citep{baugh96,kauffmann97,cole00}.
The LF in infrared bands provides the best constraints as it reliably
reflects the underlying stellar mass and is poorly sensitive to
extinction and bursts of star formation \citep{kauffmann98}.
Comparison with both the optical and infrared LFs provides tight
constraints on the models for galaxy formation \citep{baugh02}.

With the goal to derive observational measures of the galaxy LF, a
wide variety of redshift surveys with photometry from the UV to the
infrared have been analyzed. The galaxy LF is however best known in
the optical, where a wealth of details is measured.  The optical
``general'' LFs show variations from survey to survey
\citep{efstathiou88a,marzke94a,loveday95,ellis96,lin96,
zucca97,marzke98,ratcliffe98a}, which can be partly explained by the
different selection criteria used in each survey.  The large
statistical samples provided by the optical redshift surveys have also
allowed one to separately measure the LFs for different galaxy
populations, and have revealed marked differences
\citep{efstathiou88a,loveday92,marzke94b,lilly95,lin96,heyl97,lin97,
lin99,small97b,zucca97,bromley98,marzke98,metcalfe98,folkes99,loveday99,
marinoni99,brown01,fried01,madgwick02a}.

In parallel, studies of local galaxy concentrations have provided
detailed understanding of the galaxy LF, by showing that each
morphological type has a distinct LF, denoted ``intrinsic'' LF, with
different parametric functions for the giant and the dwarf galaxies
\citep[see the review by][]{binggeli88}.  \citet{sandage85b},
\citet{ferguson91}, and \citet{jerjen97b} show that the giant galaxies
have Gaussian LFs, with the LF for Elliptical galaxies skewed towards faint
magnitudes; in contrast, the LFs for dwarf galaxies may be ever
increasing at faint magnitudes to the limit of the existing surveys,
with a steeper increase for the dwarf Elliptical galaxies (dE)
compared to the dwarf Irregular galaxies (dI). 

Despite a widely varying behavior of the intrinsic LFs at faint
magnitudes for the different galaxy types, they conspire to produce in
most redshift surveys a ``general'' LF with a flat, or nearly flat
shape at faint magnitudes \citep[see for
example][]{geller97,loveday92}. Interpretation of the ``general'' LF
is complex because it results from the combination of the intrinsic
LFs with the relative proportion of galaxies in each galaxy class and
in the various environments probed by the survey. For example, the
relative density of giant galaxy types is a function of galaxy density
(as measured in clusters and groups by the morphology-density
relation, \citealp{dressler80,postman84}). Density-dependent effects
are also present in the dwarf galaxy LFs
\citep{binggeli90,ferguson91,trentham02a}. And the redshift surveys
are known to probe regions with widely varying densities, like voids,
groups, clusters, etc... \citep[see][]{lapparent86,ramella90}.  It is
therefore difficult to derive reliable constraints on the intrinsic LF
for any given galaxy type from the sole knowledge of the ``general''
LF.  As emphasized by \citet{binggeli88}, a complete characterization
of the ``general'' galaxy LF requires measurement of the intrinsic LFs
for each galaxy population.

The key for a robust measure of the intrinsic LFs is to reliably
separate the different galaxy morphological types. To this end, most
of the redshift surveys have been submitted to some galaxy
classification scheme.  The ``nearby'' redshift surveys ($z\la 0.1$)
are based on photographic catalogues, for which visual morphological
classification has been obtained
\citep{efstathiou88a,loveday92,marzke94b,marzke98,marinoni99}. These
surveys however do not explicitly include the low surface brightness
dSph (for dwarf Spheroidal, comprising dE and dS0) and dI galaxies
detected in the surveys of local galaxy concentrations
\citep{sandage85b,ferguson91,jerjen97b}. The recent morphological
analysis of a sub-sample of the Sloan Digital Sky Survey
\citep{nakamura03} based on CCD imaging to $z\sim0.1$ however shows
evidence for a contribution from dwarf galaxies.  At redshifts larger
than $\sim0.1$, visual morphological classification becomes highly
uncertain and is replaced by spectral classification
\citep{heyl97,bromley98,lin99,folkes99,fried01,madgwick02a,wolf03,lapparent03a}.
Other redshifts surveys for which a spectral classification is not
available use either colors \citep{lilly95,lin97,metcalfe98,brown01}
or the strength of the emission lines
\citep{lin96,small97b,zucca97,loveday99} for estimating the LFs of the
different galaxy types.  The widely varying criteria used for galaxy
classification in systematic redshift surveys however complicate the
interpretation and inter-comparison of the derived LFs.

In the following, I examine all the existing measurements of intrinsic
LFs obtained from optical redshift surveys at $z\la0.6$, and I convert
them into the Johnson-Cousins $UBVR_\mathrm{c}I_\mathrm{c}$ system
when other photometric systems are used. This allows a homogeneous
comparison of the intrinsic LFs measured in each band. Note that here,
the denomination ``redshift survey'' means ``systematic survey of a
region of the sky wide enough to include both cluster/group galaxies
and field galaxies, and for which estimates of redshifts are
provided''.  I therefore include the surveys by \citet{fried01} and
\citet{wolf03}, based on medium-band photometric redshifts; both
surveys provide useful estimates of intrinsic LFs, consistent with
those from the other surveys in the $B$ band, which a posteriori
justifies their inclusion into the analysis.

The article is organized as follows. In \sct \ref{LFlocal}, I recall
the properties of the intrinsic LFs based on galaxy morphological type
derived from the nearby galaxy concentrations.  In \sct \ref{LFcomp},
I analyze all existing measurements of intrinsic LFs in the $U$ and
$V$ Johnson bands (\sct \ref{UVcomp}), in the $R_\mathrm{c}$ Cousins
band (\sct \ref{Rcomp}), in the $B$ Johnson band (\sct \ref{Bcomp}),
and in the $I_\mathrm{c}$ Cousins band (\sct \ref{Icomp}). \sct
\ref{genLF} comments on the relation between the intrinsic LFs and the
``general'' LF. Finally, \sct \ref{concl} summarizes the salient
results and discusses the prospects raised by the present analysis.

\section{The local luminosity functions per morphological type         \label{LFlocal}}

Throughout the following sections, I use the estimated shape of the
intrinsic LFs per galaxy morphological type to interpret the measured
LFs from redshift surveys. Such a comparison has the advantage to
provide clues on the morphological types included in the various
classes of the considered samples.  \citet{lapparent03a} have first
emphasized the interest of this approach. The authors show that the
$R_\mathrm{c}$ LFs for the early, intermediate and late spectral
classes of the ESO-Sculptor redshift survey can be successfully
modeled as composites of the LFs measured locally for the known
morphological types of giant and dwarf galaxies. The success in using
this approach for a redshift survey prompts to extend it to the
general comparison performed here.

Following the seminal paper by \citet{sandage85b}, which shows that
the LFs of Elliptical, Lenticular and Spiral galaxies in the Virgo
cluster are bounded at bright \emph{and} faint magnitudes,
\citet{jerjen97b} derive from the joint analysis of the Virgo,
Centaurus, and Fornax clusters a robust determination of the
parametric forms for the intrinsic LFs of giant galaxies: the LFs for
S0 and Spiral galaxies have Gaussian shapes, and the LF for Elliptical
galaxies is well fitted by a two-wing Gaussian (a Gaussian with two
different dispersion wings at the bright and faint end), which is
skewed towards fainter magnitudes. In contrast, the LFs for dwarf
Spheroidal galaxies (denoted dSph) and dwarf Irregular galaxies
(denoted dI) are well fit by Schechter functions.  The dSph LF has an
ever increasing LF at the faint end, whose slope depends on the local
galaxy density
(\citealp{sandage85b,ferguson91,pritchet99,jerjen00,flint01a,flint01b,
conselice02})\footnote{Background galaxies may however contaminate the
measurements \citep[see][]{kambas00,valotto01,hilker03}.}, whereas
the dI LF appears to decrease at the faintest magnitudes with a poorly
determined shape \citep{ferguson89a,jerjen97b,jerjen00}. Moreover, in
all cases examined, the faint end of the LF for the dI galaxies
appears to be flatter than for the dSph galaxies \citep{pritchet99}.

\begin{table*}
\caption{Estimated parameters of the local Gaussian and Schechter LFs for the
different morphological types, in the Johnson-Cousins $B$, $V$, and  $R_\mathrm{c}$ bands.}
\label{LFlocal_tab}
\begin{center}
\begin{tabular}{lcccclll}
\hline 
\hline 
Morph{.} type &\multicolumn{3}{c}{Gaussian $M_0 - 5 \log h$} & Gaussian $\Sigma$ & \multicolumn{3}{c}{$\phi_0$ $^{\mathrm{a}}$} \\
              & $R_\mathrm{c}$ $^{\mathrm{b}}$  & $V$ $^{\mathrm{b}}$ & $B$ $^{\mathrm{b}}$ &   & $R_\mathrm{c}$  & $V$  & $B$ \\
\hline
E             & $-20.0\pm0.4$ & $-19.4\pm0.4$ & $-18.4\pm0.4$ & $2.1\pm0.4$, $1.3\pm0.2$ $^{\mathrm{c}}$ & 0.00046 & 0.00043 & 0.00042\\
S0            & $-20.5\pm0.1$ & $-19.9\pm0.1$ & $-19.1\pm0.1$ & $1.1\pm0.1$           & 0.00130  & 0.00126  & 0.00118  \\
\hline
Sa/Sb         & $-21.2\pm0.2$ & $-20.7\pm0.2$ & $-19.9\pm0.2$ & $0.9\pm0.1$           & 0.00699  & 0.00702  & 0.00727  \\
Sc            & $-19.8\pm0.2$ & $-19.3\pm0.2$ & $-18.7\pm0.2$ & $1.2\pm0.1$           & 0.00515  & 0.00670  & 0.00800  \\
Sd/Sm         & $-17.7\pm0.2$ & $-17.4\pm0.2$ & $-17.1\pm0.2$ & $0.8\pm0.1$           & 0.00417  & 0.00542  & 0.00648  \\
\hline
              &\multicolumn{3}{c}{Schechter $M^* - 5 \log h$} & Schechter $\alpha$  & \multicolumn{3}{c}{$\phi^*$ $^{\mathrm{a}}$} \\  
              & $R_\mathrm{c}$ $^{\mathrm{d}}$ & $V$ $^{\mathrm{d}}$ & $B$ $^{\mathrm{d}}$ &   & $R_\mathrm{c}$  & $V$  & $B$   \\
\hline
dSph          & $-18.9\pm0.3$ & $-18.4\pm0.3$ & $-17.6\pm0.3$ & $-1.7\pm0.6$ / $-1.3\pm0.1$ $^{\mathrm{e}}$ & 0.007 & 0.007 & 0.007 \\
\hline
dI            & $-17.7\pm0.3$ & $-17.4\pm0.3$ & $-17.1\pm0.3$ & $-1.3\pm0.8$ / $-0.3\pm0.2$ $^{\mathrm{e}}$ & 0.04  & 0.05  & 0.06  \\
\hline
\end{tabular}
\smallskip
\\
\end{center}
\begin{list}{}{}
\item[\underbar{Table notes}:]
\item[$^{\mathrm{a}}$] The amplitudes $\phi_0$ and $\phi^*$ are given in \phiunitn, and are derived from \protect\citet[][see text for details]{lapparent03a}.
\item[$^{\mathrm{b}}$] From \protect\citet{sandage85b} and
\protect\citet{jerjen97b} for E, S0, dSph, and dI galaxies;
estimated from \protect\citet{sandage85b} by
\protect\citet{lapparent03a} for Sa/Sb, Sc, Sd/Sm galaxies. All values
of $M_0$ and $M^*$ are converted into the $BVR_\mathrm{c}$ system by
\protect\citet{lapparent03a}.
\item[$^{\mathrm{c}}$] For E galaxies, the values of $\Sigma_a$,
$\Sigma_b$ \resp are listed (see \eq \protect\ref{gaussian_2wing}).
\item[$^{\mathrm{d}}$] For dSph and dI galaxies, the values of $M^*$
estimated from the ESO-Sculptor Survey are listed (\protect\citealp{lapparent03a}; see text for details)
\item[$^{\mathrm{e}}$] For dSph and dI galaxies, the values of $\alpha$
derived from both the Centaurus and Virgo clusters \resp are listed \protect\citep{jerjen97b}.
\end{list}
\end{table*}

\begin{figure*}
  \resizebox{\hsize}{!}
    {\includegraphics{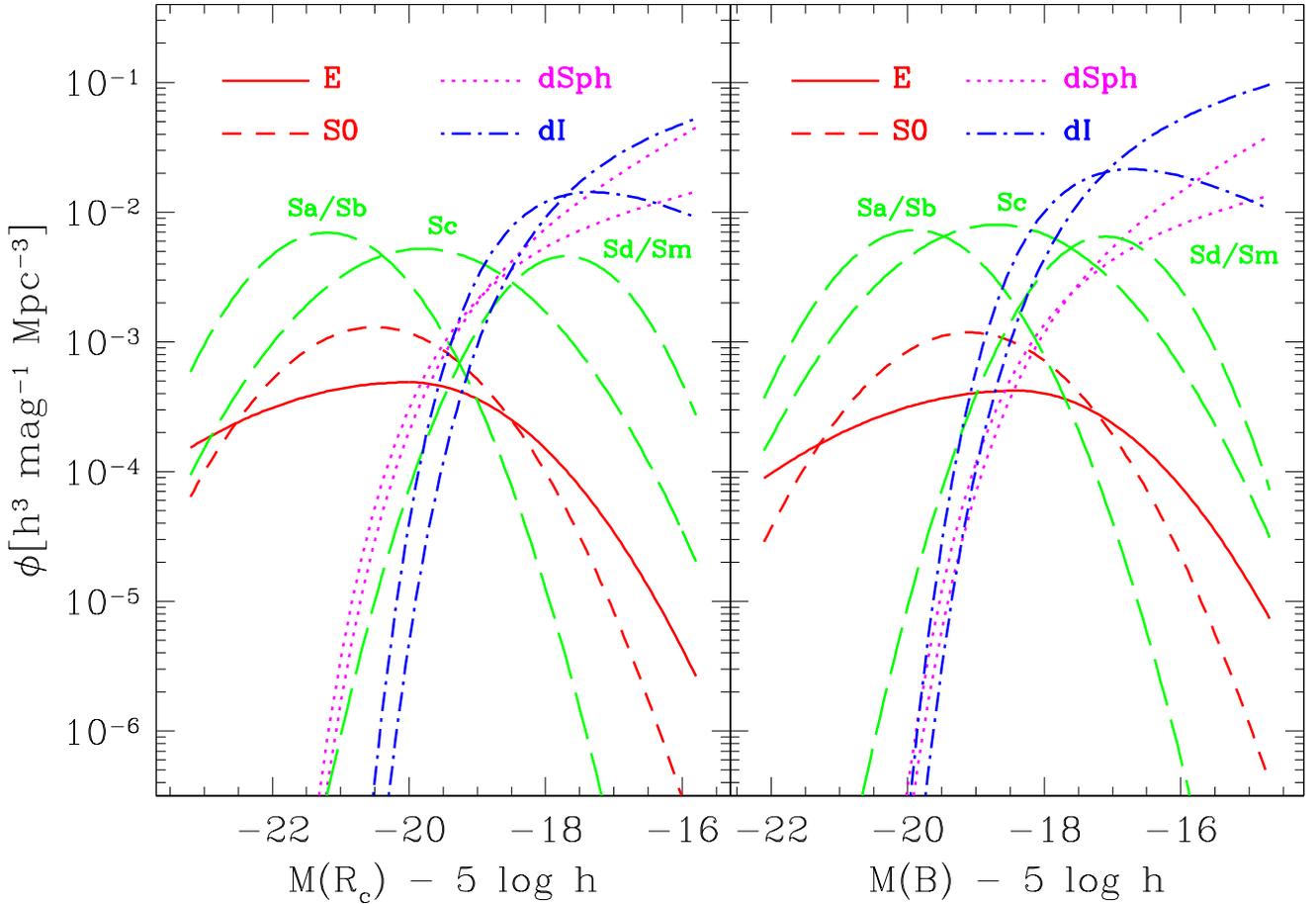}}
\caption{Intrinsic LFs in the $R_\mathrm{c}$ (left panel) and $B$
(right panel) filters with the parameters listed in Table
\ref{LFlocal_tab} for the morphological types E, S0, Sa/Sb, Sc, Sd/Sm,
dSph, and dI. The magnitude scale in the $B$ band is shifted by $1.1
^\mathrm{mag}$ (color of an Sbc galaxy, \protect\citealt{fukugita95})
towards fainter galaxies compared to the $R_\mathrm{c}$ band. To
describe the range of slopes $\alpha$ measured for the dSph and dI,
the 2 LFs with the values of $\alpha$ measured in Virgo and Centaurus
\resp are plotted for each type (see Table \ref{LFlocal_tab}).  The
amplitudes of the LFs are chosen using the results of
\protect\citet{jerjen97b} and \protect\citet{lapparent03a}. This graph
shows the relative contribution to the general luminosity functions
from the various morphological types, as a function of absolute
magnitude and filter.}
\label{LFlocal_fig}
\end{figure*} 

Table \ref{LFlocal_tab} lists the shape parameters measured by
\citet{jerjen97b} for the two-wing Gaussian and pure Gaussian LFs of
early-type giant galaxies (E, S0) in the $B_\mathrm{T}$ system, and
the converted values into $BVR_\mathrm{c}$ Johnson-Cousins magnitudes
by \citet{lapparent03a} using the results of \citet{schroeder96a} and
\citet{fukugita95}.  The Gaussian LF is parameterized as
\begin{equation}
\label{gaussian}
\phi(M) dM = \phi_0 e^{-( M_0 - M ) ^2 / 2 \Sigma^2}\; dM,
\end{equation}
where $M_0$ and $\Sigma$ are the peak and \rms dispersion respectively.
Similarly, the two-wing Gaussian is parameterized as
\begin{equation}
\label{gaussian_2wing}
\begin{array}{ll}
\phi(M) dM & = \phi_0 e^{-( M_0 - M ) ^2 / 2 \Sigma_a^2}\; dM \; {\rm for}\; M\le M_0\\
           & = \phi_0 e^{-( M_0 - M ) ^2 / 2 \Sigma_b^2}\; dM \; {\rm for}\; M\ge M_0\\
\end{array}
\end{equation}
As \citet{jerjen97b} do not provide the uncertainty in $M_0$ for the
Elliptical galaxies, I adopt a conservative error of $0.4 ^\mathrm{mag}$
(assuming a similar ratio of the uncertainty in $M_0$ by the
uncertainty in $\Sigma$ as for the S0 galaxies).  Table
\ref{LFlocal_tab} also lists the Gaussian parameters for the
individual Spiral types Sa/Sb, Sc, Sd/Sm, estimated by
\citet{lapparent03a} from \citet{sandage85b}.

Nearby redshift surveys indicate that $\sim30-40$\% of the total
number of galaxies in a redshift survey is expected to lie in groups
\citep{ramella02}, the rest lying in the so-called ``field''.
Ideally, one should therefore compare the intrinsic LFs from redshift
surveys to those derived from both field and group of galaxies.
\citet{binggeli90} do derive LFs for the different types of galaxies
in the Ursa Major Cloud (see their \fg 10), but the statistic is too
low to derive usable parameterized LFs from these data. The lack of
measurements of the intrinsic LFs for \emph{field} galaxies with a
statistical quality comparable to those for \emph{groups/clusters} of
\citet{sandage85b} and \citet{jerjen97b} leads me to refer principally
to the latter for defining the shape of the intrinsic LFs listed in
Table \ref{LFlocal_tab}.  I nevertheless refer to the field+group LFs
derived from the ESO-Sculptor Survey by \citet{lapparent03a} for
obtaining estimates of: (i) the characteristic magnitude $M^*$ for the
dSph and dI LFs; (ii) the amplitudes of the each intrinsic LF listed
in Table \ref{LFlocal_tab} (see below).

For the dwarf galaxies, the \citet{schechter76} parameterization
of the LF is
\begin{equation}
\begin{array}{ll}
\label{schechter_mag}
\phi(M) dM & = 0.4 \ln 10\; \phi^* e^{-X} X^{\alpha+1}\; dM \\
{\rm with}& \\
X & \equiv {L\over L^*} = 10^{\;0.4\,(M^* - M)}, \\
\end{array}
\end{equation}
where $M^*$ is the characteristic magnitude, and $\alpha$ the
``faint-end slope''. The values of $M^*$ for the dSph and dI LFs
listed in Table \ref{LFlocal_tab} are those estimated from the
ESO-Sculptor Survey in the $R_\mathrm{c}$ band \citep[][see their
Table 7; I use the average $M^*$ for the $R_\mathrm{c}\le20.5$ and
$R_\mathrm{c}\le21.5$ LFs, and the comparable uncertainties obtained
from the 2 measurements]{lapparent03a}. The values of $M^*$ in the
$R_\mathrm{c}$ band are then converted into the $B$ and $V$ bands
using the colors of Sab and Im galaxies: $B-R_\mathrm{c}=1.34$,
$V-R_\mathrm{c}=0.56$; $B-R_\mathrm{c}=0.58$, $V-R_\mathrm{c}=0.31$,
\resp \citep[][see their Table 3a]{fukugita95}. In the $R_\mathrm{c}$
filter, the value of $M^*$ for the dSph LF estimated from the
ESO-Sculptor Survey \citep{lapparent03a} is $\sim 0.5 ^\mathrm{mag}$
fainter than in the Virgo cluster \citep{jerjen97b}, which in turn is
$\sim 0.9 ^\mathrm{mag}$ fainter than the value measured in the
Centaurus cluster \citep{jerjen97b}. For the dI galaxies, the value of
$M^*(R_\mathrm{c})$ estimated from the ESO-Sculptor Survey
\citep{lapparent03a} is intermediate between those measured from the
Centaurus and Virgo clusters \citep{jerjen97b}.\footnote{The
ESO-Sculptor Survey also excludes the faint value of $M^*$ measured
from the Virgo cluster \citep[see][]{lapparent03a}.}

In contrast, I list for the dSph and dI LFs in Table \ref{LFlocal_tab}
the Schechter slope $\alpha$ measured from both the Centaurus and
Virgo clusters (\citealp{sandage85b}; \citealp{jerjen97b}; the steeper
slopes correspond to the Centaurus cluster), as these pairs of values
describe the range of results obtained for the dSph and dI LFs \resp
from the concentrations of galaxies of varying richness; they also
include those derived by \citet{lapparent03a} from the ESO-Sculptor
Survey.  Note that the listed parameters for the dSph and dI LFs where
derived by \citet{sandage85b} and \citet{jerjen97b} from dE+dS0, and
dI+BCD (for ``Blue Compact Galaxy'') respectively; the dE and dI
galaxies however largely outnumber the dS0 and BCD galaxies \respn, in
both the Virgo and Centaurus clusters.

All values of the LF shape parameters ($M_0$, $\Sigma$, $M^*$,
$\alpha$) listed in Table \ref {LFlocal_tab} are rounded up/down to
the first decimal place. Moreover, the listed uncertainties for the
giant galaxy types are those provided for the $B_\mathrm{T}$
measurements of the LFs by \citet{sandage85b} and
\citet{jerjen97b}. One should a priori increase the uncertainties when
performing the conversion into the $BVR_\mathrm{c}$ bands. However,
the uncertainties in the LF parameters are only listed here as
indicative of the accuracy of the quoted measurements, which frees me
from a more detailed treatment.

For graphical comparison of the intrinsic LFs listed in Table
\ref{LFlocal_tab}, one needs to define their respective amplitude.
For the dwarf LFs, I adopt and list in Table \ref{LFlocal_tab}
the average between the values measured from the ESO-Sculptor
$R_\mathrm{c}\le20.5$ and $R_\mathrm{c}\le21.5$ samples:
$\phi^*(R_\mathrm{c})=0.007$\phiunit for the dSph LF and
$\phi^*(R_\mathrm{c})=0.04$\phiunit for the dI LF \citep[see Table
7][]{lapparent03a}; the large uncertainties in these estimates lead us
to use only 1 significant digit.

To determine the amplitude $\phi_0(R_\mathrm{c})$ of the Gaussian LFs
for the giant galaxy classes listed in Table 1, I use a combination of
constraints derived in the $R_\mathrm{c}$ band from the ESO-Sculptor
Survey and the Virgo and Centaurus clusters (note that when the Virgo
and Centaurus LFs provide different constraints, I favor the Centaurus
cluster as its lower spatial density better reflects the density of
the numerous galaxy groups present in a redshift survey;
\citealt{ramella02}).  The upper bound of all the integrals mentioned
below are obtained by converting the $M(B_\mathrm{T})\simeq-15.5$
completeness limit from \citet{jerjen97b} into the $R_\mathrm{c}$ band
using the quoted colors extracted from Table 5 of
\citet{lapparent03a}.

For the individual Spiral classes, I use the following constraints,
which fully determine the values of $\phi_0(R_\mathrm{c})$ for the
Sa/Sb, Sc, and Sd/Sm LFs:
\begin{itemize}
\item the integral to $M(R_\mathrm{c})\le-16.8$ of the intrinsic LF
for Sc galaxies is equal to the integral to $M(R_\mathrm{c})\le-16.8$
of the intrinsic LF for the Sa/Sb galaxies (the color of an Sbc
galaxies is used in both cases), and is twice the integral to
$M(R_\mathrm{c})\le-16.6$ for the intrinsic LF for Sd/Sm galaxies (the
color of an Scd galaxies is used), as suggested by the results from
the Centaurus cluster \citep[see \fg 3 of][]{jerjen97b}; in the Virgo
cluster, the same ratio of the integrals of the Sc and Sd/Sm LFs is
observed whereas the ratio of Sa/Sb to Sc galaxies is only $\sim1/2$
\citep[see \fg 3 of][]{jerjen97b}. The constraint from the Centaurus
cluster provides a relative normalization for the amplitudes $\phi_0$
of the Sa/Sb, Sc, and Sd/Sm LFs;
\item the sum of half the LF for Sa/Sb galaxies and half the LF for Sc
galaxies peaks at $\phi(M)=0.005$\phiunitn, as obtained for the
Gaussian component adjusted to the ESO-Sculptor intermediate-type LF
(see Table 7 and \fg 11 of \citealt{lapparent03a}; I assume that half
of the Sa/Sb galaxies contribute to each of the ESO-Sculptor early and
intermediate-type LFs, and half of the Sc galaxies contribute to each
of the ESO-Sculptor intermediate and late-type LFs). Combined with the
preceding constraints, this provides the absolute amplitudes $\phi_0$
for the Sa/Sb, Sc and Sd/Sm LFs.
\end{itemize}

The amplitudes $\phi_0$ of the E and S0 LFs are obtained using the
following constraints:
\begin{itemize}
\item the integral to $M(R_\mathrm{c})\la-17.15$ of the intrinsic LF
for E galaxies is a factor of 2 smaller than the integral to
$M(R_\mathrm{c})\la-17.15$ of the LF for S0 galaxies (the average
color between those for an E and S0 galaxy is used), which is an
acceptable approximation of the results for both the Centaurus and
Virgo clusters \citep[see \fg 3 of][]{jerjen97b};
\item the sum of the E, S0 and half the Sa/Sb LFs peaks at
0.005\phiunitn, as estimated by the two-wing Gaussian fitted to the
ESO-Sculptor early-type LF (see Table 7 and \fg 11 of
\citealp{lapparent03a}; as already said, I assume that half the Sa/Sb
galaxies contribute to each of the ESO-Sculptor early and
intermediate-type LFs).
\end{itemize}

The resulting amplitude $\phi_0$ in the $R_\mathrm{c}$ band for each
giant and dwarf galaxy type is listed in Table \ref{LFlocal_tab}.  For
all morphological types, I then convert the values of
$\phi_0(R_\mathrm{c})$ and $\phi^*(R_\mathrm{c})$ into the $V$ and $B$
bands by multiplying by the ratio of amplitudes
$\phi^*(V)/\phi^*(R_\mathrm{c})$ and $\phi^*(B)/\phi^*(R_\mathrm{c})$
\respn, measured from the Schechter fits to the ESO-Sculptor
spectral-type LFs \citep[][see their Table 3]{lapparent03a}: for the E
and S0 LFs listed in Table \ref{LFlocal_tab}, I use the amplitude
ratios for the early-type LFs; for the Sa/Sb, and dSph LFs, those for
the intermediate-type LFs; for the Sc, Sd/Sm, dI LF, those for the
late-type LFs. I emphasize that the resulting values of $\phi_0$ and
$\phi^*$ in the $R_\mathrm{c}$,$V$, and $B$ bands listed in Table
\ref{LFlocal_tab}, are only intended as indicative of the proportions
of galaxy types expected in a redshift survey with similar selection
effects as in the ESO-Sculptor Survey.

The intrinsic LFs in the $R_\mathrm{c}$ and $B$ bands for the
parameters listed in Table \ref{LFlocal_tab} are plotted in \fg
\ref{LFlocal_fig}. The 2 graphs illustrate how each morphological type
contributes to the ``general'' LF in each band.  Any measure of LF for
a given galaxy sub-sample extracted from a redshift survey is then
expected to be some linear combination of the various LFs plotted in
\fg \ref{LFlocal_tab}, determined by the morphological content of
the sample. A wide variety of LF shapes are therefore expected, in
agreement with the diversity of results obtained from the surveys
described in the following \sctsnp.

More specifically, \fg \ref{LFlocal_fig} indicates that in both the
$R_\mathrm{c}$ and $B$ filters, the bright-end the LF for a redshift
survey is systematically dominated by one or several classes among E,
S0, Sa/Sb and Sc galaxies, depending of the galaxy types contained in
the analyzed sub-sample. For early-type galaxies (based for example on
spectral classification or colors), Sc galaxies will poorly contribute
to the LF bright-end, whereas for a late-type LF, they might fully
determine it.  Figure \ref{LFlocal_fig} also shows that in both the
$R_\mathrm{c}$ and $B$ filters, the faint-end of the LF for a
sub-sample of galaxies with intermediate spectral type or color, and
reaching a lower surface brightness than that for typical giant
galaxies, might have its faint-end dominated by dSph galaxies. The
faint-end of the LF for the bluest or latest-type galaxies is also
expected to have a contribution from the dI galaxies, which might
dominate over the Spiral galaxy types.

Note that in \fg \ref{LFlocal_fig}, the plotted magnitude interval in
the $B$ band is shifted by $1.1 ^\mathrm{mag}$ towards fainter
galaxies compared to that in the $R_\mathrm{c}$ band; this shift
corresponds to the color of an Sbc galaxy \citep[see Table 3a
of][]{fukugita95}.  Both panels of \fg \ref{LFlocal_tab} are therefore
nearly centered on the Gaussian LF for Sc galaxies.  The main
differences between the $R_\mathrm{c}$ and $B$ band are then caused by
both the $B-R_\mathrm{c}$ colors of the different galaxy types
relative to an Sc galaxy, and the variations in the LF amplitudes with
filter.  Whereas the galaxy colors are intrinsic
\citep[see][]{fukugita95}, the LF amplitudes $\phi_0$ and $\phi^*$
result from the choice which I make of the conversion factors from the
$R_\mathrm{c}$ into the $B$ band (see above).  From the $R_\mathrm{c}$
to the $B$ band, the color effects relative to the Sc LF are a dimming
of the E, S0, Sa/Sb and dSph LFs, and a brightening of the Sd/Sm and
dI LFs; the amplitude effects are an increased relative contribution
of the Sc, Sd/Sm and dI galaxies compared to the E, S0, Sa/Sb and dSph
galaxies.

\section{Comparison of the luminosity functions from redshift surveys  \label{LFcomp}}

All the existing LFs per galaxy class measured from redshift surveys
have been fitted by a Schechter (1976) function, characterized by an
exponential decrease at bright magnitudes and a power-law behavior at
faint magnitudes (see \eq \ref{schechter_mag}).  When the Schechter
parameterization LF is written as a function of absolute magnitude, as
in \eq \ref{schechter_mag} above, and is viewed in logarithmic
coordinates, the faint end has a linear behavior, with a slope
$\alpha+1$.  The value $\alpha=-1$ is therefore commonly referred to
as a ``flat slope''. As shown in the following \sctsn, the faint end
of the LF in the different surveys describes all possibilities from a
steep decrease to a flat or steep increase, which can be modeled by
varying values of the Schechter ``slope'' $\alpha$. Moreover, the
value of $M^*$ constrains the location of the exponential fall-off of
the Schechter function at bright magnitudes. The shapes of the
intrinsic LFs can therefore be conveniently compared among them using
only the values of $M^*$ and $\alpha$ of the Schechter
parameterization. 

The $M^*$ and $\alpha$ parameters for each surveys are listed in
Tables \ref{mstar_alpha_UV_tab}, \ref{mstar_alpha_R_tab}, and
\ref{mstar_alpha_B_tab}. Figures \ref{mstar_alpha_UV},
\ref{mstar_alpha_R}, and \ref{mstar_alpha_B} provide graphical
comparisons of the listed values as a function of filter and redshift
interval.  For a consistent comparison of the intrinsic LFs for the
various surveys, the values of $M^*$ in Tables
\ref{mstar_alpha_UV_tab}--\ref{mstar_alpha_B_tab} and \fgs
\ref{mstar_alpha_UV}, \ref{mstar_alpha_R}, and \ref{mstar_alpha_B}
have been converted into the $UBVR_\mathrm{c}I_\mathrm{c}$
Johnson-Cousins system; the color corrections are indicated in the
following \sctsnp. This conversion has only been performed for the
surveys in which the filter listed in \col 3 of Tables
\ref{mstar_alpha_UV_tab}--\ref{mstar_alpha_B_tab} is not among the
$UBVR_\mathrm{c}I_\mathrm{c}$ filters.

For most of the surveys considered here, the values of $M^*$ and
$\alpha$ were originally derived with $H_0 = 100 h$ km s$^{-1}$
Mpc$^{-1}$, $\Omega_m=1.0$, and $\Omega_\Lambda=0.0$.  The few surveys
for which the LFs were only measured for $\Omega_m=0.3$ and
$\Omega_\Lambda=0.7$ (the surveys denoted CS, COMBO-17 and
SDSS-Morph), have been converted into $\Omega_m=1.0$ and
$\Omega_\Lambda=0.0$ (see \sct \ref{Ucomp}). In the text, when
referring to a value of $M^*$, I omit the term $+5 \log h$, assumed to
be implicit (see Tables \ref{mstar_alpha_UV_tab} to
\ref{mstar_alpha_B_tab}).  Moreover, the uncertainties in $M^*$ and
$\alpha$ provided by all authors in the original filters are kept
unchanged when converting into the Johnson-Cousins system (except for
$\alpha[B]$ of the SDSS, see \eq \ref{sdss}). For simplicity and
because the error ellipses are not provided by the authors, I plot the
$\pm$1-$\sigma$ error-bars in $M^*$ and $\alpha$.  There is
nevertheless a correlation between the 2 parameters
\citep{schechter76} which makes a joint increase or decrease of $M^*$
and $\alpha$ less significant than an increase in $M^*$ and a decrease
in $\alpha$ by the same amount (and vice-versa). When required, the
effects of this correlation are taken into account throughout the
present analysis.

When provided by the authors, various other parameters defining each
sample from which the LFs were calculated are listed in Tables
\ref{mstar_alpha_UV_tab} to \ref{mstar_alpha_B_tab}: survey name,
solid angle, filter in which the LFs were originally calculated,
limiting magnitude of the redshift sample, effective depth or redshift
interval used in the calculation of the LF, galaxy class, number of
galaxies.  The ``effective depth'' $z_\mathrm{max}$ of a survey is
defined here as the redshift of an $M^*$ galaxy at the apparent
magnitude limit of the sample. I however have not calculated this
value for each survey, and sometimes only provide an estimate which
might differ from the true effective depth by $\la 20$\%. When
available, the actual redshift interval over which an LF is calculated
is listed instead of the effective depth.

In \fgs \ref{mstar_alpha_UV}, \ref{mstar_alpha_R}, and
\ref{mstar_alpha_B}, the points for a given survey are connected from
one class to the next, starting with the earliest class and in order
of later type.  When the intrinsic LFs are based on galaxy spectral
types, the [$M^*$,$\alpha$] points for that survey are connected by a
solid line. For LFs based on morphological types, the connecting line
is dotted. For a criterion based on color or the equivalent width of
emission lines, the connecting line is dashed.

\subsection{$U$ and $V$ bands    \label{UVcomp}}

\begin{figure*}
  \resizebox{\hsize}{!}
    {\includegraphics{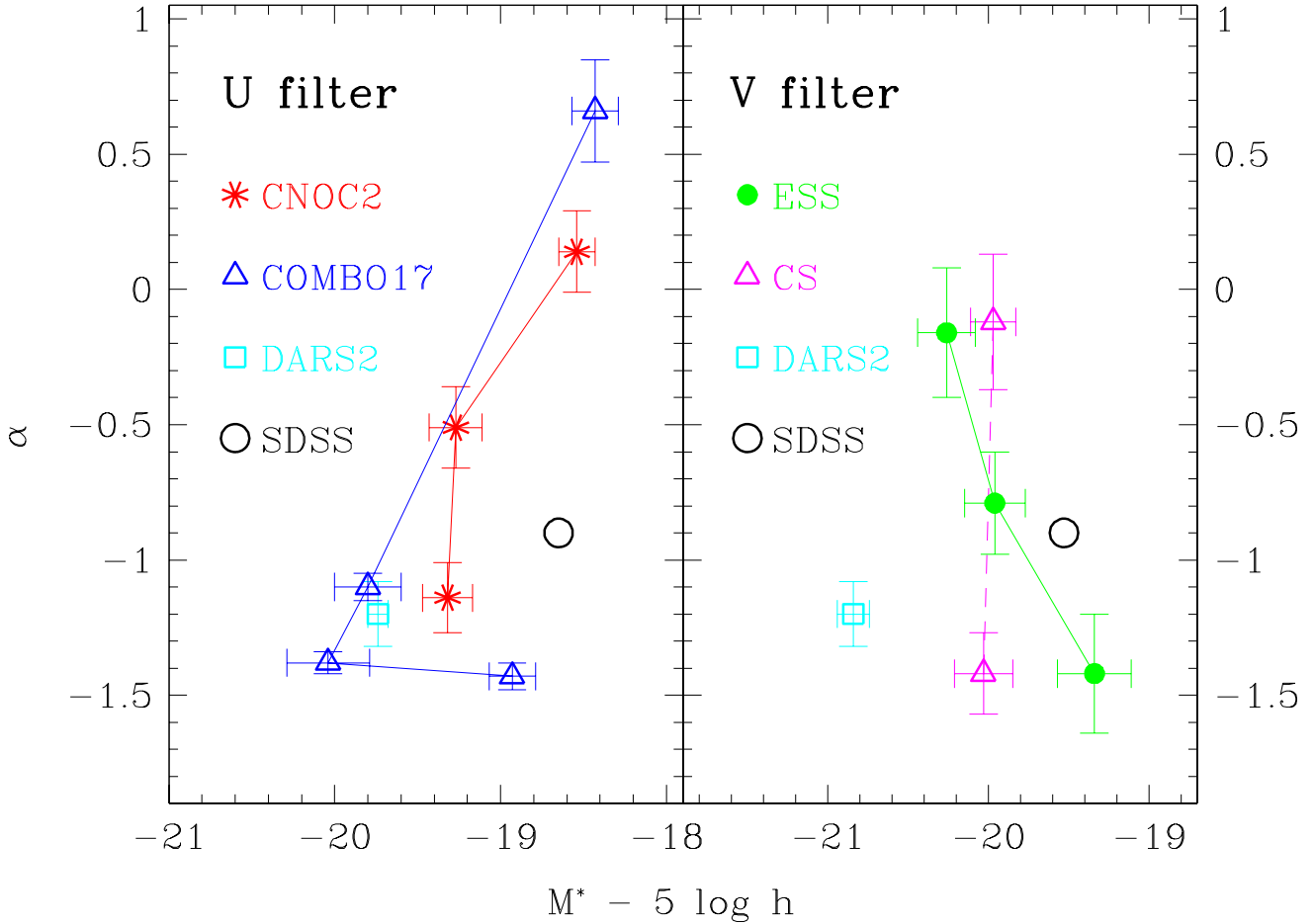}}
\caption{Comparison of the Schechter parameters $M^*$ and $\alpha$ for
the existing intrinsic LFs measured or converted into the Johnson $U$
and $V$ bands (see Table \ref{mstar_alpha_UV_tab} for the survey
parameters). Other existing surveys providing \emph{only} a general LF
are also indicated (DARS2, SDSS); as the error bars
for the SDSS survey are smaller than the symbol size (see Table
\ref{mstar_alpha_UV_tab}), they are not plotted. Solid, dashed lines
connect the various classes of a given survey when these are based on
spectral types, or a color cut, respectively. For all surveys,
galaxies of later type or with bluer colors are in the direction of
steeper slopes $\alpha$ (towards negative values).}
\label{mstar_alpha_UV}
\end{figure*}

The fewest measures of the galaxy intrinsic LFs among the
Johnson-Cousins optical bands were obtained in the $U$ and $V$
filters. The values of $M^*$ and $\alpha$ in both bands are plotted in
\fg \ref{mstar_alpha_UV} and are listed in Table
\ref{mstar_alpha_UV_tab}, along with the survey parameters.

\begin{table*}
\caption{Schechter parameters for the $U$ and $V$ intrinsic or general LFs measured from the existing redshift surveys.}
\label{mstar_alpha_UV_tab}
\begin{center}
\begin{tabular}{lrllrlrccl}
\hline
\hline
Survey        & Area & $\lambda$ & $m_\mathrm{lim}$ & $z$ & Class   &$N_\mathrm{gal}$ 
& $M^* - 5 \log h$ & $\alpha$ & Comment \\
$\;\;$ (1)   & (2) & (3) & (4) & (5) & (6) & (7) & (8) & (9) & $\;\;\;\;\;$(10) \\
\hline
DARS2   &    70.3  & $U$      & $B\le 17.0$           & 0.06       & ALL           &  288 & $-19.74\pm 0.06$ & $ -1.20\pm 0.12$ & $\alpha$ fixed from $B$ LF\\
SDSS    & $\sim 2000$   & $u^*$    & $18.36$                & 0.02--0.14 & ALL           & 22020 & $-18.65\pm 0.04$ & $ -0.90\pm 0.06$ & \\
COMBO-17$^{\mathrm{a}}$  &  0.78  & $m_{280}$ & $R \la24.0$ & 0.2--0.4   & Type-1        &  344 & $-18.43\pm 0.14$ & $\;\;\;0.66\pm 0.19$   & fits of obs. SEDs \\
         &  0.78  & $m_{280}$ & $R\la24.0$ & 0.2--0.4   & Type-2        &  986 & $-19.80\pm 0.20$ & $-1.10\pm 0.05$ & of redshifted temp.\\
         &  0.78  & $m_{280}$ & $R\la24.0$ & 0.2--0.4   & Type-3        & 1398 & $-20.04\pm 0.25$ & $-1.38\pm 0.04$ & \\
         &  0.78  & $m_{280}$ & $R\la24.0$ & 0.2--0.4   & Type-4        & 2946 & $-18.93\pm 0.14$ & $-1.43\pm 0.05$ & \\
CNOC2   &  0.692   & $U$      & $R_\mathrm{c} < 21.5$ & 0.12--0.55 & Early         &  611 & $-18.54\pm 0.11$ & $\;\;\; 0.14\pm 0.15$ & least-square fit of obs{.}  \\ 
        &  0.692   & $U$      & $R_\mathrm{c} < 21.5$ & 0.12--0.55 & Interm        &  518 & $-19.27\pm 0.16$ & $ -0.51\pm 0.15$ & $UB_\mathrm{AB}VR_CI_C$ colors \\
        &  0.692   & $U$      & $R_\mathrm{c} < 21.5$ & 0.12--0.55 & Late          & 1017 & $-19.32\pm 0.15$ & $ -1.14\pm 0.13$ & to redshifted temp{.}\\
\hline
CS$^{\mathrm{a}}$ &    65.3  & $V$    & 16.7  & 0.1        & 1/3-red    &  415 & $-19.97\pm 0.14$ & $-0.12\pm 0.25$ & $(V-R)_\mathrm{rest}>0.551$ \\
                       &    65.3  & $V$    & 16.7  & 0.1        & 1/3-blue   &  424 & $-20.03\pm 0.18$ & $-1.42\pm 0.15$ & $(V-R)_\mathrm{rest}<0.494$ \\
SDSS                   & $\sim 2000$  & $g^*$  & 17.69 & 0.02--0.17 & ALL        &  53999 & $-19.53\pm 0.02$ & $-0.91\pm 0.03$ & \\
ESS                    &   0.245  & $V$    & 21.0  & 0.1--0.6   & Early  &  156 & $-20.26\pm 0.18$ & $-0.16\pm 0.24$ &PCA-spectral class{.} \\
                       &   0.245  & $V$    & 21.0  & 0.1--0.6   & Interm &  169 & $-19.96\pm 0.19$ & $-0.79\pm 0.19$ & \\
                       &   0.245  & $V$    & 21.0  & 0.1--0.6   & Late   &  168 & $-19.34\pm 0.23$ & $-1.42\pm 0.22$ & \\
\hline
\end{tabular}
\smallskip
\\
\end{center}
\begin{list}{}{}
\item[\underbar{Table notes}:]
\item[-] Wherever necessary, the listed values of $M^*$ result from the 
         conversion from the original values derived by the authors in the
         filters listed in column (3), into the Johnson $U$ and $V$ bands, 
         respectively. The original values of $\alpha$ are kept unchanged.
\item[-] All references are provided in the text.
\item[$^{\mathrm{a}}$] The values of $M^*$ and $\alpha$ for the CS and 
COMBO-17 surveys are converted from a cosmology with [$\Omega_m=0.3$,
$\Omega_\lambda=0.7$] into [$\Omega_m=1.0$, $\Omega_\lambda=0.0$]
using the empirical corrections described in the text. These values should
therefore be used with caution.
\item[Table \cols are:]
\item[(1)] Name of survey.
\item[(2)] Survey area in square degrees.
\item[(3)] Filter in which the intrinsic or general LFs were
originally calculated by the authors.
\item[(4)] Limiting magnitude of the photometric sample, in the filter
given in the \col (3) by default, or in some other specified filter.
\item[(5)] If one value is given, it is the estimated effective depth
$z_\mathrm{max}$ of the survey (see text for details), or an upper
redshift cut-off. If an interval is given, it is the actual redshift
interval used for calculation of the corresponding LF.
\item[(6)] Galaxy class defining the sub-sample used for calculation
of the corresponding intrinsic LF.  When based on morphological types,
classes are referred to by the Hubble type.  ``Early'', ``interm'',
``late'', ``Type-'' and ``Clan-'' refer to spectral types.  ``ALL''
indicates that the ``general'' LF is listed; this is used for the
samples for which no intrinsic LFs are provided: SDSS, DARS2, and
DUKST (see \sct \ref{genLF} for details).  The scheme for galaxy 
classification is specified in the last column labeled ``Comment''.
\item[(7)] Number of galaxies in the sample/sub-sample used for calculation of the LF.
\item[(8)] Characteristic magnitude of the LF Schechter parameterization 
for the sample/sub-sample.
\item[(9)] Slope at faint magnitudes of the LF Schechter parameterization 
for the sample/sub-sample.
\item[(10)] Comment on the sample/sub-sample, plus specification of
the scheme used for classifying galaxies: if too
long, the description of the classification scheme is written over
several rows of the table; it however applies to all classes of the
considered survey.
\end{list}
\end{table*}

\subsubsection{$U$ band    \label{Ucomp}}

In the Johnson $U$ band, there is only \emph{one} survey providing
measurements of intrinsic LFs: the CNOC2 (Canadian Network for
Observational Cosmology) redshift survey \citep{lin99}, plotted in the
left panel of \fg \ref{mstar_alpha_UV}. The spectral classification
for the CNOC2 galaxies is based on least-square fits of the observed
$UBVR_\mathrm{c}I\mathrm{c}$ colors to those computed from the galaxy
spectral energy distributions (SEDs hereafter) of the templates by
\citet{coleman80}; the 4 types used in these fits are E, Sbc, Scd, and
Im, which define ``early'' (E), ``intermediate'' (Sbc), and ``late''
(Scd+Im) spectral types. Although the CNOC2 detects evolutionary
effects in $M^*$ for the $U$ intrinsic LFs \citep{lin99}, I only
consider here the $U$ LFs defined by the listed value of $M^*$ at
$z=0.3$; as no evolution is detected in $\alpha$, I use the unique
value provided by the authors. Note that $z=0.3$ corresponds
approximately to both the median and peak redshift of the survey
\citep[see \fg 6 of][]{lin99}.

Following the general trend detected in most surveys and in all
optical bands (see next \sctsn), the faint-end slope $\alpha$ for the
CNOC2 $U$ LFs becomes steeper for later type galaxies. However, in
contrast to the general dimming of $M^*$ for later galaxy types seen
in most surveys in the $BVR_\mathrm{c}$ bands (see next \sctsn), the
values of $M^*$ for the intermediate-type and late-type CNOC2 LFs are
similar, and are also $\sim0.7 ^\mathrm{mag}$ brighter than for the
early-type LF. Because $-1\la \alpha\la0$ for the 3 CNOC2
spectral-type samples, the differences in $M^*$ are a good measure of
the relative shift in the bright-end of the corresponding LFs. This
shift can be explained by the increasing emission in the UV due to
star formation in Spiral and Irregular galaxies galaxies \citep[see
for example][]{treyer98}, making the $U$ magnitude a biased estimate
of the total mass of the galaxies.

Recent estimates of UV intrinsic LFs are also provided by the COMBO-17
survey \citep[for ``Classifying Objects by Medium-Band Observations in
17 Filters''][]{wolf03}, in a synthetic UV continuum band at $\sim
2800$\AA, denoted $m_{280}$. In contrast to a redshift survey such as
the CNOC2, the spectroscopic catalogue for the COMBO-17 survey is
based on a combination of 5 wide-band filters (Johnson $UBVRI$) and 12
medium-band filters (with FWHM $\simeq140-310$\AA): spectral types and
``photometric'' redshifts are obtained by maximizing the summed
probability that an observed spectrum matches each template of a given
class among the spectral library of \citet{kinney96}; to this purpose,
the authors have replaced the noisy regions of the \citet{kinney96}
spectra using the PEGASE templates \citep{fioc97}.  Four spectral
classes are then defined: Type-1, Type-2, Type-3 and Type-4,
corresponding to Kinney et al{.} templates E-Sa, Sa-Sbc, Sbc-SB6, and
SB6-SB1 \resp (SB is ``Starburst'', and the number is defined by the
value of the $E(B-V)$ color excess, with larger numbers for increasing
color excess).  The resulting redshift uncertainties are
$\sigma(z)\le0.03$, to be compared with $\sigma(z)\sim0.0003$ for the
CNOC2.  Although the COMBO-17 LFs are derived for increasing intervals
of redshift up to $z\sim1.2$, here I only examine the intrinsic LFs in
the interval $0.2 \la z \la 0.4$, as it corresponds to the same median
redshift ($z\simeq0.3$) as in the CNOC2.

The COMBO-17 LFs are only provided for $\Omega_m=0.3$ and
$\Omega_\Lambda=0.7$. To convert to $\Omega_m=1.0$ and
$\Omega_\Lambda=0.0$, I define empirical corrections as follows.  When
changing from [$\Omega_m=0.3$, $\Omega_\Lambda=0.7$] to
[$\Omega_m=1.0$, $\Omega_\Lambda=0.0$], the variation in absolute
magnitude due to the change in luminosity distance is $\Delta
M\simeq0.3 ^\mathrm{mag}$ at $z\simeq0.3$. A change in $M^*$ of $0.3
^\mathrm{mag}$ is therefore expected for the 4 COMBO-17 LFs calculated
for $0.2 \la z \la 0.4$. This empirical correction is confirmed by the
results from the CADIS \citep{fried01}, based on medium-band
photometry as the COMBO-17, and described in \sct \ref{Bcomp_high}
below: the $B$ LFs for the 3 CADIS spectral types in the interval
$0.3< z < 0.5$ do show a dimming of $M^*$ by $0.3 ^\mathrm{mag}$ from
[$\Omega_m=0.3$, $\Omega_\Lambda=0.7$] to [$\Omega_m=1.0$,
$\Omega_\Lambda=0.0$].  I also compare the values of the Schechter
parameters for the high signal-to-noise LFs derived in both
Cosmologies from the SDSS \citep[][see \sct
\ref{genLF}]{blanton01}. These confirm that when changing from one
Cosmology to the other, the dimming in $M^*$ is related to the mean
variation in luminosity distance over the redshift interval described
by each sample \citep[see Tables 1 and 2 from][]{blanton01}. Moreover,
the $\Delta M$ dimming in $M^*$ is accompanied by a flattening in
$\alpha$ of $\sim\Delta M/3$, due to the strong correlation between
the 2 Schechter parameters \citep{blanton01}. I therefore convert the
$U$ COMBO-17 LFs in the interval $0.2<z<0.4$ from [$\Omega_m=0.3$,
$\Omega_\Lambda=0.7$] to [$\Omega_m=1.0$, $\Omega_\Lambda=0.0$] by
shifting $M^*$ and $\alpha$ by $+0.3 ^\mathrm{mag}$ and $+0.1$
respectively.  The COMBO-17 ``cosmology-shifted'' values of $M^*$ in
the $m_{280}$ band is then converted into the Johnson $U$ band using
the values of $m_{280}-U$ provided by \citet{wolfp}: 0.92, 0.52,
-0.09, -0.16 for the Type-1, Type-2, Type-3, and Type-4 LFs
respectively.  The resulting LF parameters are listed in Table
\ref{mstar_alpha_UV_tab}.  Note that the same ``cosmology-shift'' is
applied to the COMBO-17 LFs in the $R_\mathrm{c}$ and $B$ bands (see
\scts \ref{Rcomp_high} and \ref{Bcomp_high}).

The Johnson $U$ LFs derived from the COMBO-17 survey show similarities
and differences with the CNOC2 $U$ LFs.  Despite the
``cosmology-shift'' and the large $m_{280}-U$ color correction for the
COMBO-17 Type-1 galaxies (see above), their value of $M^*$ is
consistent with that for the early-type CNOC2 galaxies at less than the
1-$\sigma$ level (the variance in the difference between 2 measures of
$M^*$ or $\alpha$ is estimated as the quadrature sum of the
uncertainties in the 2 measures). In contrast, the COMBO-17 value of
$\alpha$ departs from the CNOC2 value by 2.1-$\sigma$. This difference
in the value of $\alpha$ is consistent with the small fraction of
galaxies included in the COMBO-17 Type-1 class (corresponding to
\citealt{kinney96} spectral types E-Sa): they represent only 6\% of
the galaxies with $0.2<z<0.4$, whereas the early-type classes in the
CNOC2 (corresponding to \citealt{coleman80} spectral type E) contains
29\% of the galaxies. Although the Type-1 galaxies in the COMBO-17 are
selected using E-Sa template spectra, this class is dominated by E and
S0 \citep[see also \fg 2 from][]{wolf03}, whereas the CNOC2 early-type
classes contain in addition a significant number of Sa and Sab
galaxies. As shown by \citet{lapparent03a}, dSph galaxies which might
cause the flat faint-end slope of the ESO-Sculptor intermediate class
can have optical colors comparable to those for Sab galaxies (see also
\sct \ref{LFlocal}); these objects might therefore also contaminate
the CNOC2 early-type LF, while being excluded from the COMBO-17 Type-1
class, which could in turn explain the ``flatter'' value of $\alpha$
for the CNOC2 early-type LF.

The $U$ LFs for both the COMBO-17 and CNOC2 show a brightening of
$M^*$ from the early to intermediate spectral classes. This
brightening is however larger for the COMBO-17, with a $1.4
^\mathrm{mag}$ brightening of $M^*$ between the Type-1 and the Type-2
LFs; it corresponds to a similar shift of the LF bright-end \citep[see
\fgs A.11 and A.12 in][]{wolf03}, to be compared to the $\sim0.5
^\mathrm{mag}$ brightening in the value of $M^*$ between the CNOC2
early-type LF and both the intermediate and late-type LFs.  The
COMBO-17 Type-3 class (\citealt{kinney96} spectral types Sbc-SB6) and
the CNOC2 intermediate-type class (\citealt{coleman80} spectral type
Sbc) are expected to have a significant number of galaxies in common,
due to their similar spectral content. The COMBO-17 Type-3 LF however
has a steeper $\alpha$ at the $5.8$-$\sigma$ level, and a brighter
$M^*$ at the $2.6$-$\sigma$ level compared to the CNOC2
intermediate-type LF. The COMBO-17 Type-2 LF (spectral types Sa-Sbc)
also has a steeper $\alpha$ at the $3.7$-$\sigma$ level and brighter
$M^*$ at the $2.1$-$\sigma$ level compared to the CNOC2 intermediate-type
LF, whereas one would expect the Type-2 LF to be intermediate between
the CNOC2 early-type (spectral types E) and intermediate-type
(spectral type Sbc).  Note that the elongation of the error ellipses
for the Schechter parameterization in the direction of brighter $M^*$
and steeper $\alpha$ would actually decrease the quoted significance
levels. These would however remain larger than $2$-$\sigma$ for the
difference in $\alpha$.

One explanation could be related to the uncertainties in the absolute
magnitudes induced by the $\sigma(z)\sim0.03$ redshift errors in the
COMBO-17. For faint Starburst galaxies, the redshifts errors are even
larger, $\sigma(z)\sim0.1$, and imply magnitude errors of $0.75$
mag. \citet[][see the end of their \sct 3.5]{wolf03} warn that these
uncertainties tend to ``bias the steep luminosity function of
Starburst galaxies to brighter $L^*$ values''.  A significant
contamination of the Type-2 (spectral types Sa-Sbc) and Type-3
(spectral types Sbc-SB6) classes by Starburst galaxies, despite the
small expected number of such objects in these classes, could explain
the bright $M^*$ for the corresponding LFs. Surprisingly, this
luminosity bias does not appear to affect the Type-4 LF (spectral
types SB6-SB1) which only differs from the CNOC2 late-type LF
(spectral types Scd/Im) at the $\sim2$-$\sigma$ level in both $M^*$ and
$\alpha$, with a fainter $M^*$ and a steeper $\alpha$; this is in
agreement with the similar fractions of galaxies in the COMBO-17
Type-4 and the CNOC2 late-type samples (52\% and 47\% \respn). The
Type-4 class is however supposed to contain \emph{only} Starburst
galaxies, for which the luminosity bias is expected to be the largest.
Other complex selection effects inherent to surveys based on
multi-medium-band photometry, and most critical for emission-line
galaxies, might also be at play in the COMBO-17. Another possible
interpretation is that the LFs for the COMBO-17 Spiral galaxies (Sa,
Sb, Sc) which dominate the Type-2 and Type-3 class and may have
significant emission-lines, may also be biased towards bright values of
$M^*$, whereas the Type-4 class succeeds in separating the lower-mass
Irregular galaxies populating the LF (see \fg \ref{LFlocal_fig}).  The
absence of systematic brightening of $M^*$ for the COMBO-17 Type-4 LF
compared to the CNOC2 late-type LF could then result from the
combination of a systematic brightening affecting the Type-4 galaxies
which would be compensated for by an intrinsic fainter $M^*$ than in
the CNOC2 late-type LF.  At last, the COMBO-17 color transformations
from the synthetic UV continuum band $m_{280}$ into the Johnson $U$
band \citep{wolfp} might suffer some biases, possibly related to the
large difference ($\sim1000$\AA) between the peaks in the response
curves of the respective filters.

\subsubsection{$V$ band    \label{Vcomp}}

There are so far only 2 estimates of intrinsic LFs in the Johnson $V$
band: the Century Survey \citep[denoted CS]{brown01}, for which the
LFs are calculated from the 1/3 blue and 1/3 red portions of the full
sample, based on $V-R_\mathrm{c}$ rest-frame color (see Table
\ref{mstar_alpha_UV_tab} for the color bounds); and the ESO-Sculptor
survey \citep[denoted ESS]{lapparent03a}, which provides the
\emph{first} measurements of $V$ intrinsic LFs based on 3 spectral
classes. LF measurements from both surveys are plotted in the right
panel of \fg \ref{mstar_alpha_UV}. Note that the ESS only detects
evolution in the amplitude of the late-type LF \citep{lapparent03b};
the listed values of $M^*$ and $\alpha$ in Table
\ref{mstar_alpha_UV_tab} are those derived from the full redshift
range of the ESS.  As the intrinsic LFs for the CS 1/3-red and
1/3-blue sub-samples are only provided for cosmological parameters
$\Omega_m=0.3$ and $\Omega_\lambda=0.7$, I convert the $M^*$, $\alpha$
values provided by the authors to $\Omega_m=1.0$ and
$\Omega_\Lambda=0.0$ by adding $0.1$ and $0.03$ \respn, based on the
variations for the full $V$ sample (see Table 2 of \citealp{brown01});
note that these offsets in $M^*$ and $\alpha$ are consistent with
those which would be inferred by the empirical method which I use
above for converting the COMBO-17 LFs to
[$\Omega_m=1.0$, $\Omega_\Lambda=0.0$], and which is based on the mean
variation of the luminosity distance over the considered sample, when
changing the cosmological parameters.

The faint-end slope $\alpha$ for both the 1/3 red and 1/3 blue LFs in
the Century Survey are in agreement with those for the ESS early-type
and late-type LFs, respectively.  In contrast, the values of $M^*(V)$
for the Century Survey LFs are nearly equal; \fgs 15 and 16 of
\citet{brown01} however show that the bright ends of the 2 LFs differ
by $0.5 ^\mathrm{mag}$. A similar effect is present in the ESS LFs:
there is a $\sim0.9 ^\mathrm{mag}$ dimming of $M^*(V)$ between the
early-type and the late-type LFs, whereas the bright-end of the
early-type and late-type LFs are shifted by a larger amount,
$\sim1.5 ^\mathrm{mag}$.  The strong correlation between $M^*$ and
$\alpha$ in the Schechter parameterization implies that the difference
in the value of $M^*$ for 2 different LFs may not be a measure of the
actual shift in the bright-end fall-off for these 2 LFs. An exact
correspondence only occurs if the 2 LFs have the same value of
$\alpha$. The various surveys examined in the present article 
show that for $-1\la \alpha\la0$, the agreement is within $0.1$
mag. For steeper values of $\alpha$, in the interval
$-2\la\alpha\la-1$, a shift $\Delta M$ in the LF bright-end
corresponds to a change in $M^*$ by $\Delta M^*\simeq\Delta M -
f(\alpha)$, with $0.5\la f(\alpha)\la 1 ^\mathrm{mag}$.

\begin{figure}
  \resizebox{\hsize}{!}
    {\includegraphics{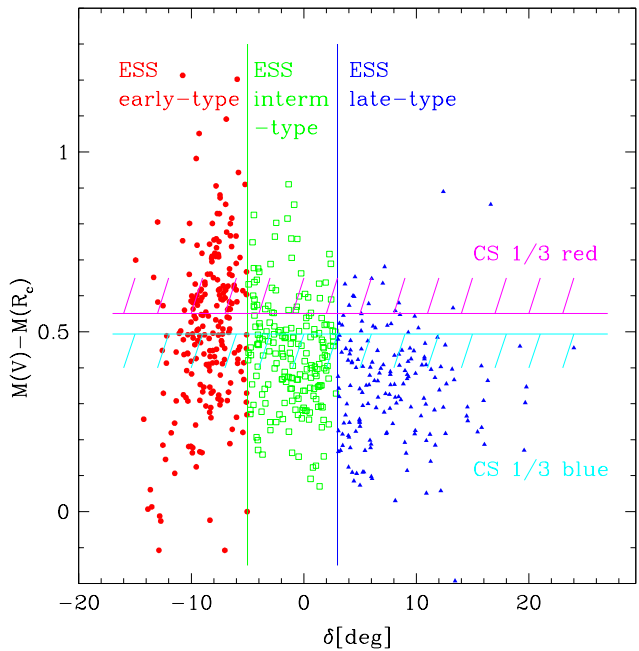}}
\caption{Relation between the ESO-Sculptor (ESS) PCA spectral type $\delta$
and the absolute color $M(V)-M(R_\mathrm{c})$ of each galaxy. The
ESO-Sculptor early, intermediate and late-type galaxies are defined by
the intervals $\delta\le-5.0^\circ$, $-5.0<\delta\le3.0^\circ$, and
$\delta>3.0^\circ$ \respn, separated by the 2 vertical
lines. The color cuts used for the measurement of the Century Survey
LFs are superimposed as the 2 horizontal lines: the Century Survey (CS) 1/3
blue and 1/3 red sub-samples are defined by
$M(V)-M(R_\mathrm{c})<0.494$ and $M(V)-M(R_\mathrm{c})>0.551$
respectively. This graph shows how sub-samples based on a color
cut mix galaxies of different spectral types.}
\label{delta_MVR}
\end{figure} 

In the ESS, the dimming of $M^*(V)$ for later spectral types confirms
that $M(V)$ is a better estimate of the total mass of the galaxies
than $M(U)$. This dimming is interpreted by \citet{lapparent03a} as a
signature of the fainter magnitude late-type Spiral galaxies (Sc, Sd,
Sm) detected in the late-type class, compared to the brighter earlier
type Spiral galaxies Sa and Sb included in the early and
intermediate-type classes \resp (see also \sct \ref{LFlocal} and \fg
\ref{LFlocal_fig}). The smaller dimming in $M^*(V)$ for the CS
compared to the ESS can be explained as a result of the color cut for
separating the LF sub-samples, which causes some mixing of the
spectral types.  Figure \ref{delta_MVR} illustrates this effect by
showing the distribution of ESS spectral types $\delta$, as a function
of absolute (or rest-frame) color $M(V)-M(R_\mathrm{c})$ for each
galaxy. In the ESS, the spectral type $\delta$ is obtained by a
classification based on a Principal Component Analysis (PCA
hereafter), and is tightly correlated with the morphological type
\citep{lapparent03a}; the early-type, intermediate-type, and late-type
classes, are defined by $\delta\le-5.0^\circ$,
$-5.0<\delta\le3.0^\circ$, and $\delta>3.0^\circ$ \resp (shown as
vertical lines in \fg \ref{delta_MVR}), and contain predominantly
E/S0/Sa, Sb/Sc, and Sc/Sm/Im galaxies \resp \citep{lapparent03a}.

By applying to the ESS the colors cuts used in the CS for defining the
1/3 red and 1/3 blue sample, \fg \ref{delta_MVR} shows that the 2
color samples contain significant fractions of galaxies from several
spectral classes: the blue sample contains $27.3$\%, $35.2$\%, and
$37.5$\% of early-type, intermediate-type, and late-type galaxies
\respn, and the red sample, $62.5$\%, $26.9$\%, and $10.6$\%
respectively.  Therefore, the red sample is dominated by the
early-type galaxies and to a smaller extent, by the intermediate-type
galaxies; in contrast, the blue sample contains comparable fractions
of the 3 galaxy spectral types. The steeper slope for the CS 1/3 blue
LF compared to the 1/3 red LF reflects the fact that the majority of
the galaxies of late spectral type are included in the 1/3 blue sample
(see \fg \ref{delta_MVR}). The 2 LFs however have comparable $M^*$
because its determination is dominated by the brightest galaxies in
the 2 color samples, namely the E/S0/Sa/Sb galaxies which populate the
early-type and intermediate-type spectral classes, both included in
the 2 color samples.  This analysis illustrates how intrinsic LFs
based on 2 color classes fail to separate the blue low luminosity
galaxies from the more luminous Elliptical and Spiral galaxies.

Note that the fainter peak surface brightness limit reached in the ESS
($\simeq 22-22.5$ $V$ mag arcsec$^{-2}$, \citealp{lapparent03a}),
compared to $\simeq 20-21$ $V$ mag arcsec$^{-2}$ in the CS
\citep{brown01},\footnote{In both surveys, the surface brightness is
corrected for the $2.5 \log(1+z)^4$ redshift-dimming; in the ESS, 
K-corrections are also applied \citep[see][]{lapparent03a}.}  might
also contribute to a better detection of Irregular galaxies which
dominate the ESS late spectral-types at faint magnitudes
($M(R_\mathrm{c})\ga-18.5$; see \fg \ref{LFlocal_fig}) and have lower
surface brightness than Elliptical and Spiral galaxies.

\subsection{$R_\mathrm{c}$ band             \label{Rcomp}}

\begin{figure*}
  \resizebox{\hsize}{!}
    {\includegraphics{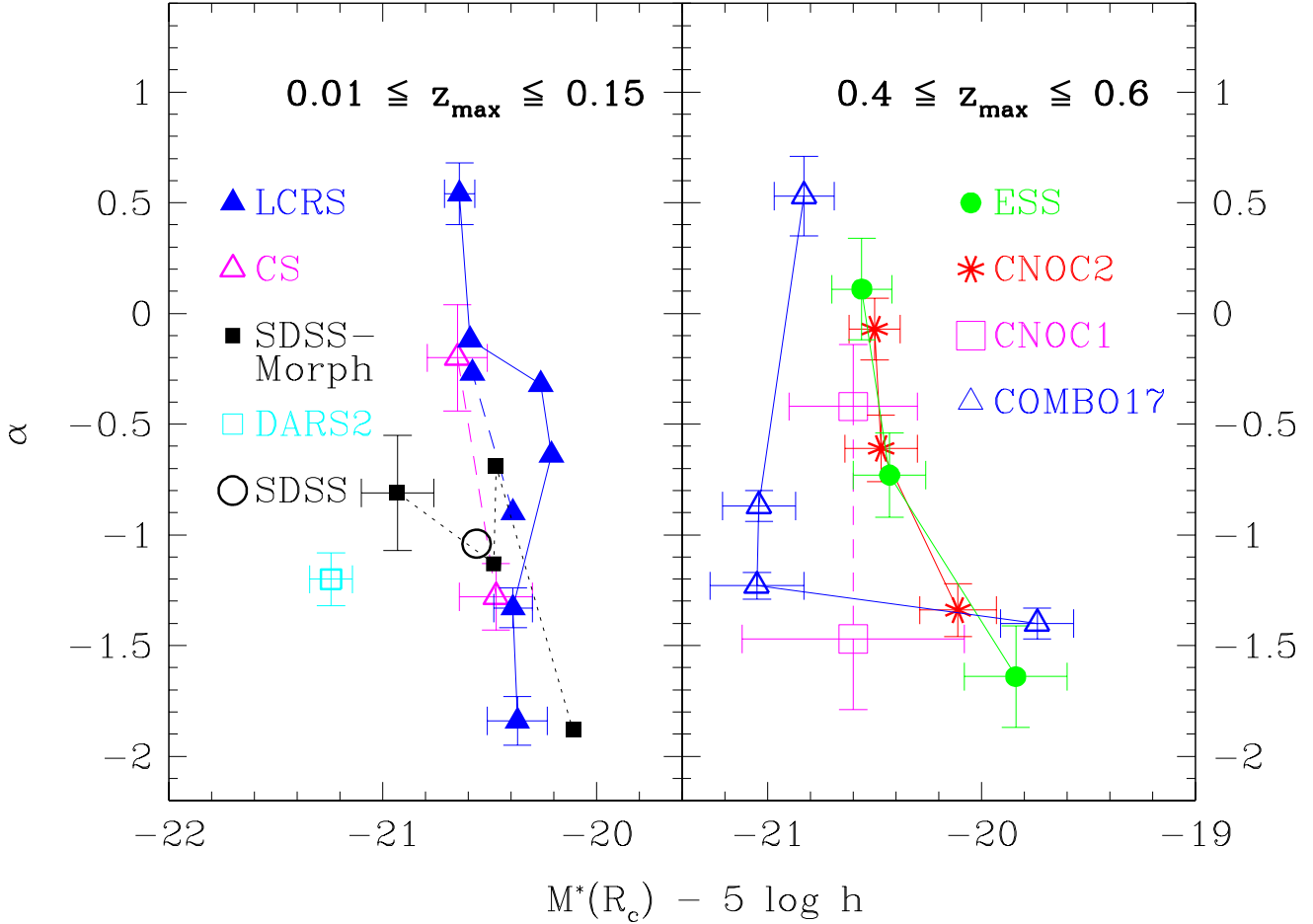}}
\caption{Comparison of the Schechter parameters $M^*$ and $\alpha$ for
the existing intrinsic LFs measured or converted into the Cousins
$R_\mathrm{c}$ band. The surveys with effective depth $0.01\le
z_\mathrm{max}\le 0.15$ are shown in the left panel, those with
$0.4\le z_\mathrm{max}\le 0.6$ in the right panel (see Table
\ref{mstar_alpha_R_tab} for the survey parameters).  The general LFs
provided by the DARS2 and SDSS are also shown (see \sct \ref{genLF};
the error bars for the SDSS are smaller than the symbol size). The error
bars are not shown for several points of the LCRS survey for which it
is nearly equal or smaller than the symbol size (see Table
\ref{mstar_alpha_R_tab}). For clarity, the error bars for the
morphological LFs of the SDSS are only shown for the E-S0 LF; the
error bars for types S0/Sa-Sb Sbc-Sd are similar, whereas no error bar
are provided for Type Im by \protect\citet{nakamura03}.  Solid,
dashed, and dotted lines connect the various classes of a given survey
when these are based on spectral types, a color cut or the equivalent
width of emission lines, and morphological types respectively. For all
surveys, galaxies of later type, with bluer colors or stronger
emission lines are in the direction of steeper slopes $\alpha$
(towards negative values). }
\label{mstar_alpha_R}
\end{figure*}

The existing measurements of intrinsic LFs in the $R_\mathrm{c}$ band
are more numerous than in the $U$ and $V$ bands. They are listed in
Table \ref{mstar_alpha_R_tab} and plotted in \fg \ref{mstar_alpha_R}.

\subsubsection{$R_\mathrm{c}$ luminosity functions at redshifts 0.4--0.6  \label{Rcomp_high}}

The right panel of \fg \ref{mstar_alpha_R} displays the estimates of
intrinsic LF from the 4 surveys with $0.4 \la z_\mathrm{max}\la0.6$:
the CNOC2 survey \citep{lin99}, already mentioned in \sct \ref{Ucomp},
and its cluster analog, the CNOC1 survey, which also provides a sample
of field galaxies \citep{lin97}; the ESS \citep{lapparent03a}, and the
COMBO-17 \citep{wolf03}, both already mentioned in \scts \ref{Vcomp}
and \ref{Ucomp} respectively. For the CNOC1, survey, the
\citet{thuan76} $r$ magnitudes are converted into $R_\mathrm{c}$ by
applying the conversion $R_\mathrm{c} = r - 0.36$, as calculated by
\citet{fukugita95} for an Sbc galaxy; no distinction is made for the
various spectral types as $r - R_\mathrm{c}$ varies in the narrow
interval $0.34-0.38$ among the 6 galaxy types listed by
\citet{fukugita95}. Because the COMBO-17 red LFs are measured in the
Sloan Digital Sky Survey $r^*$ band \citep{fukugita96}, I convert the
``cosmology-corrected'' values of $M^*(r^*)$ (see \sct \ref{Ucomp})
into the Johnson $R$ band using the values of $r^*-R$ provided by
\citet{wolfp}: 0.3, 0.3, 0.22, 0.16 for Type-1, Type-2, Type-3, and
Type-4 galaxies respectively. I then apply the $R_\mathrm{c}-R$ colors
terms provided by \citet[][see their Table 3]{fukugita95} for types
S0, Sbc, Scd, and Im: 0.09, 0.08, 0.07, and 0.03 respectively.

Because the CNOC2, COMBO-17 and ESS all detect evolutionary effects in
some of their $R_\mathrm{c}$ intrinsic LFs, I restrict the comparison
to the LFs measured at the median and/or peak redshift of
$z\simeq0.3$. As in the $U$ band, I use for the CNOC2 the listed
values of the evolving $M^*$ at $z=0.3$, and the non-evolving value of
$\alpha$ \citep{lin99}. For the COMBO-17 survey, I use the intrinsic
LFs in the interval $0.2 \la z \la 0.4$.  For the ESS, the mean LFs
over the full redshift range of the survey are provided by
\citet{lapparent03a}.

\begin{table*}
\caption{Schechter parameters for the $R_\mathrm{c}$ and $I_\mathrm{c}$ intrinsic or general LFs measured from the existing redshift surveys.}
\label{mstar_alpha_R_tab}
\begin{center}
\begin{tabular}{lrllllrccl}
\hline
\hline
Survey        & Area & $\lambda$ & $m_\mathrm{lim}$ & $z$ & Class   &$N_\mathrm{gal}$ 
& $M^* - 5 \log h$ & $\alpha$ & Comment \\
$\;\;$ (1) & (2) & (3) & (4) & (5) & (6) & (7) & (8) & (9) & $\;\;\;\;\;$(10) \\
\hline
 CS $^{\mathrm{a}}$ 
         &    65.3  & $R_\mathrm{c}$ & 16.2        & 0.1        & 1/3-red  &   419 & $-20.65\pm 0.14$ & $-0.20\pm 0.24$ & $(V-R)_\mathrm{rest}>0.555$ \\
         &    65.3  & $R_\mathrm{c}$ & 16.2        & 0.1        & 1/3-blue &   422 & $-20.47\pm 0.17$ & $-1.28\pm 0.15$ & $(V-R)_\mathrm{rest}<0.499$ \\
 DARS2   &  70.3    & $R_\mathrm{c}$ & $B\le 17.0$ & 0.06       & ALL      &   288 & $-21.24\pm 0.10$ & $-1.20\pm 0.12$ & alpha fixed from $B$ LF\\
 SDSS-Morph & 230.0 & $r^*$          & 15.90       & 0.01--0.075& E/S0     &   597 & $-20.93\pm 0.17$ & $-0.81\pm 0.26$ & $0\le T\le 1.0$  \\
         &   230.0  & $r^*$          & 15.90       & 0.01--0.075& S0/Sa/Sb &   518 & $-20.48\pm 0.19$ & $-1.13\pm 0.26$ & $1.5\le T\le 3 $ \\
         &   230.0  & $r^*$          & 15.90       & 0.01--0.075& Sbc/Sd   &   350 & $-20.47\pm 0.20$ & $-0.69\pm 0.26$ & $3.5\le T\le 5 $ \\
         &   230.0  & $r^*$          & 15.90       & 0.01--0.075& Im       &    10 & $-20.11\;\;\;\;\;\;\;\;\;\;$    & $-1.88\;\;\;\;\;\;\;\;\;\;\;$         & $5.5\le T\le 6 $ \\
 SDSS    &  $\sim 2000$  & $r^*$     & 17.79       & 0.02--0.22 & ALL      & 147986& $-20.56\pm 0.03$ & $-1.04\pm 0.03$ & \\
 LCRS    &   462    & $r$            & 17.7        & 0.15       & OII      &  7312 & $-20.03\pm 0.03$ & $-0.90\pm 0.04$ & EW[OII] $>$ 5 \AA\\
         &   462    & $r$            & 17.7        & 0.15       & no-OII   & 11366 & $-20.22\pm 0.02$ & $-0.27\pm 0.04$ & EW[OII] $<$ 5 \AA\\
         &   462    & $r$            & 17.7        & 0.15       & Clan-1   &   655 & $-20.64\pm 0.07$ & $\;\;\; 0.54\pm 0.14$ & PCA-spectral class{.} \\
         &   462    & $r$            & 17.7        & 0.15       & Clan-2   &  7614 & $-20.59\pm 0.03$ & $-0.12\pm 0.05$ & \\
         &   462    & $r$            & 17.7        & 0.15       & Clan-3   &  4667 & $-19.26\pm 0.04$ & $-0.32\pm 0.07$ & \\
         &   462    & $r$            & 17.7        & 0.15       & Clan-4   &  3210 & $-19.21\pm 0.05$ & $-0.64\pm 0.08$ & \\
         &   462    & $r$            & 17.7        & 0.15       & Clan-5   &  1443 & $-20.39\pm 0.09$ & $-1.33\pm 0.09$ & \\
         &   462    & $r$            & 17.7        & 0.15       & Clan-6   &   689 & $-20.37\pm 0.14$ & $-1.84\pm 0.11$ & \\
 COMBO-17$^{\mathrm{a}}$ &   0.78  & $r^*$          & $R\la24.0$  & 0.2--0.4   & Type-1   &   344 & $-20.83\pm 0.14$ & $\;\;\; 0.53\pm 0.18$ & fits of obs. SEDs \\
          &   0.78  & $r^*$          & $R\la24.0$  & 0.2--0.4   & Type-2   &   986 & $-21.04\pm 0.17$ & $ -0.87\pm 0.07$ & of redshifted temp. \\
          &   0.78  & $r^*$          & $R\la24.0$  & 0.2--0.4   & Type-3   &  1398 & $-21.05\pm 0.22$ & $ -1.23\pm 0.06$ & \\
          &   0.78  & $r^*$          & $R\la24.0$  & 0.2--0.4   & Type-4   &  2946 & $-19.74\pm 0.17$ & $ -1.40\pm 0.07$ & \\
 CNOC2   &  0.692   & $R_\mathrm{c}$ & 21.5        & 0.55       & Early    &   611 & $-20.50\pm 0.12$ & $-0.07\pm 0.14$ & least-square fit of obs{.} \\
         &  0.692   & $R_\mathrm{c}$ & 21.5        & 0.55       & Interm   &   517 & $-20.47\pm 0.17$ & $-0.61\pm 0.15$ & $UB_\mathrm{AB}VR_CI_C$ colors\\
         &  0.692   & $R_\mathrm{c}$ & 21.5        & 0.55       & Late     &  1012 & $-20.11\pm 0.18$ & $-1.34\pm 0.12$ & to redshifted temp{.}\\
 CNOC1   &     -    & $r$            & 22.0        & 0.2--0.6   & 1/2-red  &   209 & $-20.60\pm 0.30$ & $-0.42\pm 0.28$ & $r$-$g$ of redshifted \\
         &     -    & $r$            & 22.0        & 0.2--0.6   & 1/2-blue &   179 & $-20.60\pm 0.52$ & $-1.47\pm 0.32$ & non-evolv{.} Sbc temp{.}\\
 ESS     & 0.247    & $R_\mathrm{c}$ & 20.5        & 0.1--0.6   & Early    &   232 & $-20.56\pm 0.14$ & $\;\;\; 0.11\pm 0.23$ & PCA-spectral class{.} \\
         & 0.247    & $R_\mathrm{c}$ & 20.5        & 0.1--0.6   & Interm   &   204 & $-20.43\pm 0.17$ & $-0.73\pm 0.19$ & \\
         & 0.247    & $R_\mathrm{c}$ & 20.5        & 0.1--0.6   & Late     &   181 & $-19.84\pm 0.24$ & $-1.64\pm 0.23$ & \\
\hline
 DARS2   &    70.3  & $I_\mathrm{c}$ & $B\le 17.0$ & 0.06       & ALL      &   288 & $-21.92\pm 0.10$ & $-1.20\pm 0.12$ & alpha fixed from $B$ LF \\ 
 SDSS    & $\sim 2000$  & $i^*$      & 16.91       & 0.02--0.22 & ALL      & 88239 & $-21.25\pm 0.02$ & $-1.03\pm 0.03$ & \\
\hline
\end{tabular}
\smallskip
\\
\end{center}
\begin{list}{}{}
\item[\underbar{Table notes}:]
\item[-] See Table \ref{mstar_alpha_UV_tab} for definition of \cols
All references are provided in the text.
\item[-] Wherever necessary, the listed values of $M^*$ result from the conversion from
the original values derived by the authors in the filters listed in
column (3), into the Cousins $R_\mathrm{c}$ and $I_\mathrm{c}$ bands,
respectively. The original values of $\alpha$ are kept unchanged.
\item[-] $r$ magnitudes are in the \citet{thuan76} photometric system.
\item[$^{\mathrm{a}}$] The values of $M^*$ and $\alpha$ for the
CS, SDSS-Morph and COMBO-17 surveys are converted from
a cosmology with [$\Omega_m=0.3$, $\Omega_\lambda=0.7$] into
[$\Omega_m=1.0$, $\Omega_\lambda=0.0$] using empirical corrections
described in the text. These values should therefore be used with
caution.
\end{list}
\end{table*}

The measured values of $M^*(R_\mathrm{c})$ and $\alpha$ for the CNOC2
and ESS are in good agreement. \citet{lapparent03a} show that the ESS
$R_\mathrm{c}$ early-type LF is consistent with a Gaussian
parameterization, in agreement with the Gaussian LFs measured locally
for E, S0, and Sa galaxies (see also \fg \ref{LFlocal_fig} in \sct
\ref{LFlocal} above).  The similar $M^*$ and $\alpha$ parameters for
the CNOC2 $R_\mathrm{c}$ LF indicate that a Gaussian parameterization
might also be appropriate for the early-type LF for that sample.  Both
samples show a steepening of $\alpha$ when going to later spectral
types, and a dimming of $M^*(R)$ by $\sim0.6 ^\mathrm{mag}$ when going from
intermediate-type to late-type galaxies, with most of the dimming
occurring between the intermediate and late-type LFs. As in the $V$
band, this dimming is due to the fainter galaxies (types Sc/Sm/Im)
included in the ESS and CNOC2 late-type classes (see \fg
\ref{LFlocal_fig}).

The general agreement of the ESS and CNOC2 intrinsic LFs in the
$R_\mathrm{c}$ band is a result of the similar morphological content
of the spectral classes, as shown by \citet{lapparent03a}: the early,
intermediate, and late-type classes contain predominantly E/S0/Sa,
Sb/Sc, and Sc/Sm/Im \resp in the ESS, and E/Sab, Sbc, and Scd/Im \resp
in the CNOC2; when extrapolated to $R_\mathrm{c}\le21.5$, the ESS
early, intermediate, and late-type classes contain 27\%, 30\%, and
43\% \resp of the total number of galaxies, and the CNOC2 classes contain
29\%, 24\%, and 47\% respectively.  Given the 1-mag difference in the
magnitude limit for the 2 surveys, the differing selection effects and
redshift completeness curves, this agreement is remarkable.  The
slight shift toward galaxies of earlier type in the CNOC2 late-type
class, as indicated by the fractions of galaxies per spectral type,
might also explain why this LF has a flatter $\alpha$ and brighter
$M^*$ than in the ESS.

Comparison of the 4 spectral-type LFs for the COMBO-17 survey with the
3 spectral-type LFs for the ESS and CNOC2 yields similar conclusions
as in the $U$ band.  For the Type-1 galaxies, the value of $M^*$ is
consistent with those for the ESS and CNOC2 early-type LFs at less
than the 1-$\sigma$ level, whereas the COMBO-17 value of $\alpha$
departs from the values in the ESS and CNOC2 by 1.4-$\sigma$ and
2.6-$\sigma$ respectively. As in the $U$ band, I interpret the
systematically larger value of $\alpha$ for the COMBO-17 Type-1 LF as
due to: (i) the earlier spectral content of this class, compared to
both the ESS and CNOC2 early-type classes; (ii) the likely absence
from the COMBO-17 Type-1 class of dSph galaxies which would flatten
the faint-end of the LF (see \sct \ref{LFlocal}).

Both Schechter parameters for the COMBO-17 Type-4 LFs are in agreement
with those for the ESS and CNOC2 late-type class (at less than the
1-$\sigma$ level, except for $M^*$ which differs from that for the
CNOC2 by 1.5-$\sigma$). This suggests that there is a significant
fraction of galaxies in common between the COMBO-17 Type-4 galaxies
(with spectral types matching the Starburst templates SB6 to SB1 from
\citealt{kinney96}), and the Sc/Sm/Im and Scd/Im galaxies selected in
the ESS and CNOC2 late-type class respectively.  The fractions of
galaxies in the corresponding classes for the 3 surveys (29\% in the
ESS, 47\% in the CNOC2, and 52\% in the COMBO-17) also support a
significant common population of galaxies.

In contrast, similar differences between the COMBO-17 Type-2 and
Type-3 LFs and the CNOC2 LFs as those seen in the $U$ band are
detected in the $R_\mathrm{c}$ band.  Both Schechter parameters for
the COMBO-17 Type-3 LF (corresponding to spectral types Sbc-SB6)
significantly differ from those for the CNOC2 (Sbc) and the ESS
(Sb/Sc) intermediate-type LFs (at the $2.1$-$\sigma$ and
$2.2$-$\sigma$ level \resp for $M^*$, at the $3.8$-$\sigma$ and
$3.1$-$\sigma$ level \resp for $\alpha$), despite a significant common
spectral content (see \sct \ref{UVcomp}), with offsets in the
direction of brighter $M^*$ and steeper $\alpha$ for the COMBO-17 LFs.
As in the $U$ band, the COMBO-17 Type-2 LF (spectral types Sab-Sbc) is
expected to lie in the intermediate region between the early and
intermediate-type LF for the CNOC2 and ESS, containing E/Sab and Sbc
galaxies respectively. However, for the COMBO-17 Type-2 LF, $M^*$ is
brighter by $\sim2.4$-$\sigma$ and $\alpha$ is steeper by
$1.6$-$\sigma$ and $0.7$-$\sigma$ than for the CNOC2 and ESS
intermediate-type LFs respectively.

As already stated in \sct \ref{Ucomp}, a shift towards bright
magnitudes is expected for the COMBO-17 Starburst galaxies, and the
bright values of $M^*$ for the Type-2 and Type-3 LFs could indicate a
severe contamination of these 2 classes by Starburst galaxies.  The
low expected fraction of Starburst galaxies in these 2 classes however
suggest that a similar magnitude bias might affect the Spiral
galaxies, which dominate the Type-2 and Type-3 classes. As in the $U$
band, the similar values of $M^*$ for the COMBO-17 Type-4 LF and the
ESS and CNOC2 late-type LFs, despite the dominating fraction of
Starburst galaxies in the Type-4 class, could result from the
combination of a systematic brightening affecting the Type-4 LF
compensated for by an intrinsic fainter $M^*$ than in the ESS and
CNOC2 late-type LFs.  The complex selection effects inherent to the
use of medium-band photometry for redshift measurement and spectral
classification do not allow to discard these 2 interpretations. At
last, some systematic biases in the COMBO-17 color transformation from
the $r^*$ to the $R_\mathrm{c}$ band \citep{wolfp} might also operate,
although the difference between the response curves in the 2 filters
is significantly smaller than between the $m_{280}$ and $U$ bands (see
\sct \ref{Ucomp}).

The right panel of \fg \ref{mstar_alpha_R} also shows the CNOC1 LFs
estimated from the 1/2-red and 1/2-blue sub-samples, separated by the
redshifted $r-g$ color of an Sbc galaxy. The values of $M^*$ and
$\alpha$ for the CNOC1 1/2-red sample are intermediate between those
for the early and intermediate-type LFs for the ESS and CNOC2 samples,
suggesting an agreement with both surveys.  The 2 CNOC1 LFs also
display the steepening in $\alpha$ for bluer galaxies, to a value
comparable to those for the CNOC2 and ESS late-type LFs. Although the
CNOC1 LFs fail to detect a significant dimming in $M^*$, because of
the correlation between $M^*$ and $\alpha$, there is a
$\sim0.7 ^\mathrm{mag}$ dimming of the LF bright-end between the 2
CNOC1 samples \citep[see \fg 3 in][]{lin97}. This is to be compared to
the $\sim1.7 ^\mathrm{mag}$ dimming of the LF bright-end between the
CNOC2 and ESS early and late-type classes (\citealp[see \fg 3
in][]{lin97}, and \citealp[\fg 7 in][]{lapparent03a}).  This behavior
is similar to that for the CS 1/3 red and 1/3 blue samples in the $V$
band and illustrated in \fg \ref{delta_MVR}, and can be attributed to
same cause: the bright end of the CNOC1 1/2-red and 1/2-blue LFs are
likely dominated by Elliptical and early-type Spiral galaxies \respn,
which have similar characteristic magnitudes.

\subsubsection{$R_\mathrm{c}$ luminosity functions at redshifts 0.01--0.15  \label{Rcomp_low}}

The left panel of \fg \ref{mstar_alpha_R} gathers the few intrinsic LF
estimates from redshift surveys with $0.01 \la z_\mathrm{max}\la
0.15$.  The $R_\mathrm{c}$ LFs for the CS are calculated for the same
1/3 red and 1/3 blue sub-samples as the $V$ LFs (see Table
\ref{mstar_alpha_R_tab}); the values of $M^*$ and $\alpha$ for both
samples are empirically converted from cosmological parameters
[$\Omega_m=0.3$, $\Omega_\lambda=0.7$] into
[$\Omega_m=1.0$, $\Omega_\Lambda=0.0$] by adding $0.1$ and $0.03$
\respn, based on the variations for the full $R$ sample (see Table 2
of \citealp{brown01}; see also comments on these shifts in \sct
\ref{Ucomp}). The resulting LFs display a similar behavior to both the
CS $V$ LFs, and the CNOC1 LFs converted into the $R_\mathrm{c}$ band.
The value of $M^*$ dims by only $0.2 ^\mathrm{mag}$ from the CS 1/3 red
to the 1/3 blue $R_\mathrm{c}$ LF, corresponding to a
$0.5 ^\mathrm{mag}$ shift between the bright-end of the 2 LFs. The
values of $\alpha$ show the usual steepening from the red to blue
sample, and the values are in agreement with those for the CNOC1
(right panel of \fg \ref{mstar_alpha_R}) at less that the
\onesign.

The only measures of LF in a red filter based on morphological types
were recently obtained from a sub-sample with $r^*\le 15.9$ from the
Early Data Release (EDR) of the Sloan Digital Sky Survey
\citep[denoted here SDSS-Morph]{nakamura03}.  Following \sct
\ref{Ucomp}, I convert the listed values of $M^*(r^*)$ from
[$\Omega_m=0.3$, $\Omega_\Lambda=0.7$] to [$\Omega_m=1.0$,
$\Omega_\Lambda=0.0$] using $\Delta M\simeq0.06 ^\mathrm{mag}$, which
corresponds to the change in absolute magnitude due to the change in
luminosity distance at $z\simeq0.05$, close to the median redshift of
the sub-sample. Using the relation $\Delta\alpha\simeq\Delta M/3$,
derived from the various LFs listed in Table 2 of \citet{blanton01}, I
also apply the empirical shift $\Delta\alpha\simeq0.02$ to the
values of $\alpha$ listed by \citet{nakamura03}. For the 4 SDSS
morphological types listed in Table \ref{mstar_alpha_R_tab}, I convert
the ``cosmology-corrected'' values of $M^*(r^*)$ into the Cousins
$R_\mathrm{c}$ band using the $0.24$ average $r^*-R_\mathrm{c}$ color
over listed types E and S0, and the $0.24$, $0.23$, $0.17$ colors for
listed types Sab, Sbc, and Im respectively \citep[][in their Table
3]{fukugita95}. The resulting values of $M^*(R_\mathrm{c})$ and
$\alpha$ are listed here in Table \ref{mstar_alpha_R_tab}. Note that
the LF for types Im is only given by \citet{nakamura03} as indicative
(hence the lack of error bars), as this sample is too small and too
incomplete to provide a reliable LF.

The intrinsic LFs derived by \citet{nakamura03} show a nearly flat
slope for the 3 morphological types E/S0,S0/Sa/Sb, Sbc/Sd.  Only the
LF for morphological type Im shows a steep slope $\alpha\sim-1.9$.
This is comparable with the behavior of the morphological-type LFs
measured in the Johnson $B$ band for the NOG \citep{marinoni99}, the
CfA2S \citep{marzke94b} and the SSRS2 \citep{marzke98} surveys
(described in \sct \ref{Bcomp_low}). The flat faint-end slopes
measured by \citet{nakamura03} for types E/S0 with no evidence of a
faint-end decline is at variance with the Gaussian LFs measured
locally for E and S0 (see \sct\ref{LFlocal}). When \citet{nakamura03} use
the concentration index for classifying galaxies, they obtain a
similar flat early-type LF.  The authors interpret this flat slope as
the presence of many intrinsically faint elliptical galaxies with a
``hard core'' out to $M(r^*)\sim-19$. This is in agreement with the
fact that in the Virgo cluster, the bright-end of the dSph LF is
dominated by nucleated dE \citep[see \fgs 6 and 15
in][]{sandage85b}. Objects of this type are likely to appear as
elliptical galaxies in the visual classification by
\citet{nakamura03}.  As stated by the authors, separating the
contribution from dSph at the faint-end of the E/S0 LF might yield a
decline of this LF.

Because the surface brightness profile of dSph galaxies deviates from
the $r^{1/4}$ profile of giant E \citep{vaucouleurs48} and resembles
the exponential profile measured for disk galaxies
\citep{binggeli91,binggeli98}, I suggest that \citet{nakamura03} might
have classified some non-nucleated dE galaxies as faint Spiral
galaxies. This would explain the absence of a decline in the S0/Sa/Sb LF
at $M(r^*)$ fainter than $\sim-19$, as would be expected by
combination of the Gaussian LFs measured locally for these 3 galaxies
types (see \sct \ref{LFlocal}).  A contribution from a Schechter LF
for dSph with a steep faint-end slope $\alpha\la-1.3$ could explain
the increase of the S0/Sa/Sb LF at faint magnitude, with
$\alpha=-1.15\pm0.26$ \citep[see][]{lapparent03a}. Moreover, the faint
boundary $M(r^*)\simeq-18$ of the 3rd SDSS-Morph class
\citep{nakamura03} is too bright to show a decrease at faint
magnitudes, as this LF is expected to correspond to the combination of the Gaussian
LFs for the Sbc and Sd (see \sct \ref{LFlocal}).  Although the dwarf
Irregular galaxies (dI) mostly populate the latest class of the
SDSS-Morph sample (Im), an additional contribution at $-20\la
M(r^*)\la-18$ from dI galaxies (see Table \ref{LFlocal_tab} and \fg
\ref{LFlocal_fig} in \sct \ref{LFlocal}) might contribute to
preventing a decline of the faint-end LF for Sbc/Sd galaxies.

Contrary to the nearby surveys based on morphological types (see \sct
\ref{Bcomp_low}), the SDSS-Morph survey does detect the dimming of
$M^*$ of the Im galaxies, compared to the earlier classes.  However,
the value of $M^*(R_\mathrm{c})=-20.11$ for the Schechter Im LF is 2
to 4$ ^\mathrm{mag}$ brighter than the values measured from the Centaurus
and Virgo cluster \citep[see Table 6 of][]{lapparent03a}, and
$\sim2.5 ^\mathrm{mag}$ brighter than the value derived from the
ESO-Sculptor (see Table 7 of
\citealp{lapparent03a}). \citet{nakamura03} warn that the SDSS-Morph
Im sample is incomplete, and it is likely that a significant portion
of the ``unclassified'' objects are faint Im galaxies.  Due to the
limits of visual classification \citep{lahav95}, some type mixing
among the dSph, Im and faint Spiral galaxies, might be expected, and
could affect the various SDSS-Morph LFs.

In the left panel of \fg \ref{mstar_alpha_R}, I also plot the
$R_\mathrm{c}$ intrinsic LFs for the Las Campanas Redshift Survey
\citep[denoted LCRS]{bromley98,lin96}.  As for the CNOC1, I convert
the LCRS \citet{thuan76} $r$ magnitudes into the $R_\mathrm{c}$ band
using the $r - R_\mathrm{c} = 0.36$ color of an Sbc galaxy
\citep[][Table 3f]{fukugita95}, with no distinction of spectral type.

The LCRS intrinsic LFs based on 6 spectral classes derived by a PCA
\citep{bromley98}, show a smooth variation in $\alpha$ from $0.54$ to
$-1.84$, and a dimming of $M^*(R)$ from $-20.28$ to $-20.01$ between
Clan-1 and Clan-6.  The large value $\alpha=0.54\pm0.14$ for the
Clan-1 LF suggest that this sub-sample contains only early-type
galaxies and is not contaminated by dwarf spheroidal galaxies.
Further comparison of the LCRS LFs with those for the other surveys is
hindered by the fact that \citet{bromley98} do not provide the
correspondence between their PCA-spectral type and the Hubble
morphological types.

Moreover, the LCRS redshift survey is based on multi-fiber
spectroscopy for which a spectral classification is subject to biases
caused by:
\begin{itemize}
\item the relatively small angular size of the fibers ($3.5\arcsec$)
compared to the apparent galaxy size, which introduces systematic
color biases into the spectra (this effect is called ``aperture
bias''): color gradients are present in galaxies of varying types
\citep{segalovitz75,boroson87,vigroux88,balcells94}, and in most cases
correspond to several tenths of a magnitude bluer colors when going
from the central to the outer regions of a galaxy;
\item the astrometric uncertainties which cause an offset of the
positioned fiber with respect to the galaxy peak of light;
\item the poor flux calibration of the spectra, as the variations in
the fiber transmission cannot be accurately calibrated, implying some
dispersion in the spectra continuum shape.
\end{itemize}
To partly overcome the flux-calibration inaccuracy,
\citet{bromley98} apply to each spectrum a high-pass filter, which
effectively removes the continuum of the spectra. The PCA analysis
therefore only accounts for ``local'' features such as the CaII H\&K
break, and the absorption and emission lines. \citet{galaz98} however
show that in a spectral classification based on flux-calibrated
spectra, the dominant signal originates from the \emph{shape} of the
continuum. Moreover, the lack of accurate flux-calibration in the LCRS
also results in significant dispersion in the relative line
intensities, likely to cause some contamination among the LCRS
spectral classes (no error analysis of the random and systematic
errors in the flux-calibration of the LCRS data is however reported by
\citealt{bromley98}). The net effect is to smooth the variations among
the intrinsic LFs.  This could explain the smaller variation in $M^*$
between Clan-1 and Clan-6 ($0.27 ^\mathrm{mag}$), compared to a variation
between the early and late-type LFs of $0.62 ^\mathrm{mag}$ in the ESS,
and $0.39 ^\mathrm{mag}$ in the CNOC2. \citet{kochanek01b} also showed how
aperture biases may artificially steepen the LF by mixing galaxies
having Schechter LFs with similar faint-end slopes but different $M^*$
and different amplitudes.  The LFs derived by \citet{bromley98} must
therefore be used with caution. In \sct \ref{Bcomp_mid_high} below, I
show that the various biases mentioned here may also affect the $B$
band determinations of the LFs provided by surveys based on
multi-fiber spectroscopy.

\begin{figure}
  \resizebox{\hsize}{!}
    {\includegraphics{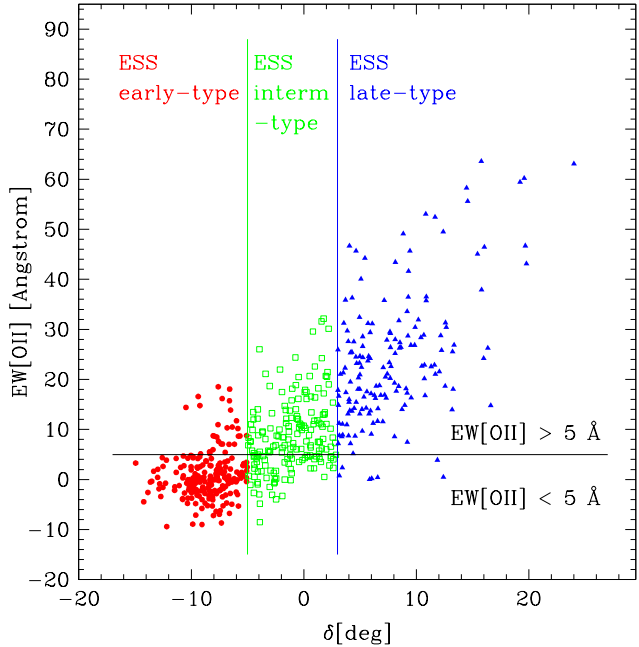}}
\caption{Relation between the ESO-Sculptor PCA spectral type $\delta$
and the equivalent width of the [OII] emission line for each
ESO-Sculptor galaxy. The ESO-Sculptor early, intermediate and
late-type galaxies are defined by the intervals $\delta\le-5.0^\circ$,
$-5.0<\delta\le3.0^\circ$, and $\delta>3.0^\circ$ \respn, separated by
the 2 vertical lines. The cut at EW[OII]=5\AA~used for measurement of
the Las Campanas Redshift Survey LFs \protect\citep{lin96} is
indicated as a horizontal line (EW[OII]=5\AA also nearly corresponds to a
2-$\sigma$~significance level in the [OII] line). This graph shows how
sub-samples based on EW[OII] mix galaxies of different spectral
types.}
\label{delta_WOII}
\end{figure} 

Left panel of \fg \ref{mstar_alpha_R} also shows the intrinsic LFs
estimated from the LCRS using the sub-samples with EW[OII] $>$ 5
\AA~and EW[OII] $<$ 5 \AA~\resp \citep{lin96}. Although the
H$\alpha$$\lambda$6563 emission line is a more reliable indicator of
star-formation than the [OII]$\lambda$3727 line as it is less affected
by dust and metallicity \citep{tresse99}, the [OII] line is often used
at $z\ga 0.3$ where the H$\alpha$ lines shifts into the infrared. The
2 LCRS LFs based on EW[OII] show the similar dimming in $M^*$ and
steepening in $\alpha$ as seen between the LFs for Clan-2 and an
intermediate LF between those for Clan-4 and Clan-5.  Emission lines
provide a convenient and straightforward method for separating
galaxies with early and late morphological types, as nebular lines
result from gas heating by young stars and are thus present in
galaxies with current star formation, which in turn tend to be of
later morphological type (see \fg \ref{delta_WOII}, described
below). Although the correlations between strength of the nebular
lines, the galaxy color/spectral-type and the morphological type
suffer some dispersion, they are observed in all galaxy samples
\citep[see \fgs 2 and 3 in][]{lapparent03a}.  For example,
\citet{heyl97} show that the evolution detected by \citet{ellis96} in
the Autofib star-forming galaxies from $z\simeq0.5$ to the present time (a
decrease in luminosity density with decreasing redshift), can be
interpreted in terms of evolution in the late-type Spiral galaxies
(see \sct \ref{Bcomp_mid_low} for analysis of the Autofib intrinsic
LFs).

I now show that LF estimates based on EW[OII] suffer analogous type
mixing as those derived from color samples (see \fg
\ref{delta_MVR}). In \fg \ref{delta_WOII}, I plot the equivalent width
of the [OII] emission line as a function of PCA spectral type $\delta$
for the galaxies with $R_\mathrm{c}\le20.5$ in the ESS. The sample
with EW[OII]$<5$\AA~contains $72.6$\%, $24.6$\%, and $2.8$\% of
early-type, intermediate-type, and late-type galaxies \respn, and the
sample with EW[OII]$>5$\AA, $9.9$\%, $38.9$\%, and $51.2$\%
respectively. Therefore, the low [OII]-emission sample is dominated by
the early-type galaxies, with a small fraction of intermediate-type
galaxies and few late-type galaxies; in contrast, the high
[OII]-emission sample is approximately equally dominated by the
intermediate and late-type galaxies.  Measurement of LFs based on the
equivalent width of [OII] emission line then fails to discriminate
among the intrinsic LFs per morphological type due to type mixing,
similarly to the LFs based on color sub-samples. This could explain
why the $R_\mathrm{c}$ LFs for the CS 1/3 red and 1/3 blue samples
nearly follow the LCRS results based on the [OII] emission line (left
panel of \fg \ref{mstar_alpha_R}).  There is however one notable
difference with the LFs obtained by using a color cut: the fraction of
early-type galaxies in the high [OII]-emission ESS sample is
relatively smaller than in the ESS-1/3 blue sample.

The LCRS LFs are also useful for emphasizing the need of multiple
galaxy classes for estimating intrinsic LFs.  The difference in the
LCRS LFs between the 6 samples separated by spectral type and the 2
samples based on the strength of the [OII] emission line illustrates
how a wider variety of LFs is measured when a larger number of classes
is used. This is due to the multiplicity of shape for the LFs per
morphological type (see \sct \ref{LFlocal}).  In the $R_\mathrm{c}$
band, comparison of the LCRS and CS LFs on one hand, and of the CNOC1
and CNOC2 LFs on the other hand, provides evidence that a minimum of 3
spectral classes is necessary for detecting both the Gaussian LF shape
for the E and S0 galaxies (sometimes also including Sa/Sb galaxies),
and the dimming of the late-type Spiral (Sc, Sd/Sm) and the Irregular
galaxies compared to earlier-type galaxies (see \fg
\ref{LFlocal_fig}).  Comparison of the ESS and CS LFs in the $V$ band
(see \fg \ref{mstar_alpha_UV} in \sct \ref{Vcomp}) also supports this
result.

\subsection{$B$ band       \label{Bcomp}}

\begin{figure*}
\resizebox{\hsize}{!}{\includegraphics{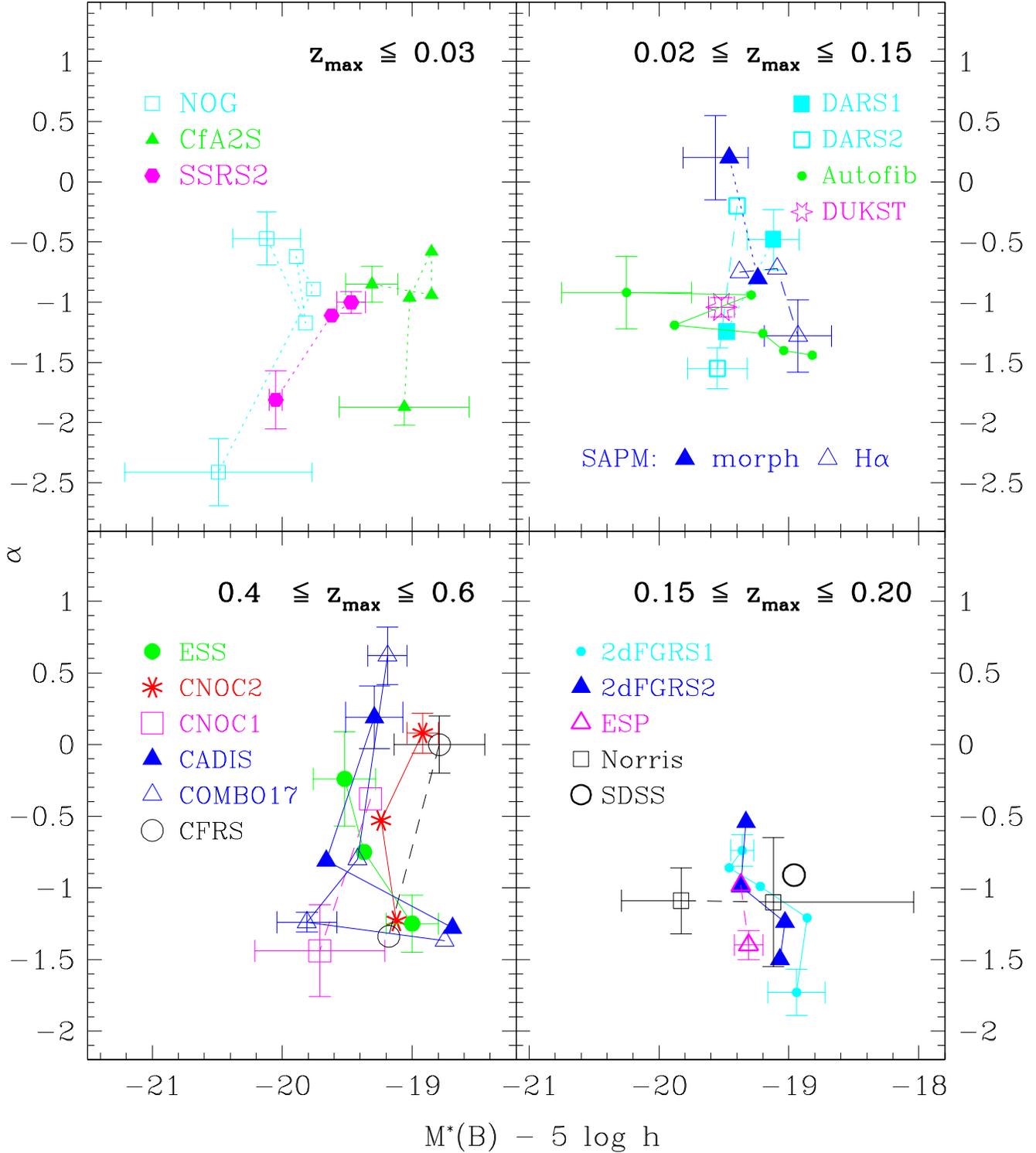}}
\caption{Comparison of the Schechter parameters $M^*$ and $\alpha$ for
the existing intrinsic LFs measured or converted into the Johnson $B$
band. Other existing surveys providing \emph{only} a general LF are
also indicated (DUKST in the upper-right panel; SDSS in the lower-right panel).
As the error bars for the 2dFGRS2 and SDSS surveys (lower-right panel)
are smaller than the symbol size (Table \ref{mstar_alpha_B_tab}), they
are not plotted. The 4 panels are arranged by increasing effective
depth $z_\mathrm{max}$, starting from the upper-left panel and moving
clock-wise; the interval of effective depth is indicated in each
panel.  Solid, dotted, and dashed lines connect the various classes of
a given survey when these are based on spectral types, morphological
types, and a color cut or the equivalent width of emission lines,
respectively.  For all surveys, galaxies of later
spectral/morphological type or with stronger emission lines are in the
direction of steeper slopes $\alpha$ (towards negative values), except
for the Norris survey in the lower-right panel, for which $\alpha$ is
nearly constant: the ``late-type'' galaxies are those with the fainter
$M^*$. For clarity, the error bars in
the lower-left panel are only shown for either the early-type point or
the late-type point of each survey, or for both points; error bars for
the other points are comparable, except in the COMBO-17, with similar
error bars for the Type-2, Type-3 and Type-4 LFs, and in the ESS,
with similar error bars for the intermediate-type and late-type LFs
(see Table \ref{mstar_alpha_B_tab}).}
\label{mstar_alpha_B}
\end{figure*} 

\begin{table*}
\caption{Schechter parameters for the $B$ intrinsic or general LFs measured from the existing redshift surveys.}
\label{mstar_alpha_B_tab}
\begin{center}
\begin{tabular}{lrlp{1.6truecm}rp{1.2truecm}rccl}
\hline
\hline
Survey        & Area & $\lambda$ & $m_\mathrm{lim}$ & $z$ & Class   &$N_\mathrm{gal}$ 
& $M^* - 5 \log h$ & $\alpha$ & Comment \\
$\;\;$ (1)   & (2) & (3) & (4) & (5) & (6) & (7) & (8) & (9) & $\;\;\;\;\;$(10) \\
\hline
 NOG     & 27140 & $B$             & 14.0                     & 0.02       & E              &  344 & $-20.12\pm 0.26$ & $-0.47\pm 0.22$ & morphological class{.} \\
         & 27140 & $B$             & 14.0                     & 0.02       & S0             &  596 & $-19.82\pm 0.26$ & $-1.17\pm 0.20$ &  \\         
         & 27140 & $B$             & 14.0                     & 0.02       & Sa-Sb          & 1521 & $-19.89\pm 0.12$ & $-0.62\pm 0.11$ &  \\ 
         & 27140 & $B$             & 14.0                     & 0.02       & Sc-Sd          & 2240 & $-19.76\pm 0.11$ & $-0.89\pm 0.10$ &  \\ 
         & 27140 & $B$             & 14.0                     & 0.02       & Sm-Im          &  619 & $-20.49\pm 0.72$ & $-2.41\pm 0.28$ &  \\ 
 CfA2S   &  1370 & $B_\mathrm{Zw}$ & 15.5                     & 0.03       & E              &  -   & $-19.31\pm 0.20$ & $-0.85\pm 0.15$ & morphological class{.} \\
         &  1370 & $B_\mathrm{Zw}$ & 15.5                     & 0.03       & S0             &  -   & $-18.85\pm 0.10$ & $-0.94\pm 0.10$ & \\ 
         &  1370 & $B_\mathrm{Zw}$ & 15.5                     & 0.03       & Sa-Sb          &  -   & $-18.85\pm 0.10$ & $-0.58\pm 0.10$ & \\ 
         &  1370 & $B_\mathrm{Zw}$ & 15.5                     & 0.03       & Sc-Sd          &  -   & $-19.02\pm 0.15$ & $-0.96\pm 0.10$ & \\ 
         &  1370 & $B_\mathrm{Zw}$ & 15.5                     & 0.03       & Sm-Im          &  -   & $-19.06\pm 0.50$ & $-1.87\pm 0.15$ & \\ 
 SSRS2   &  5550 & $B_\mathrm{Zw}$ & 15.5                     & 0.03       & E-S0           & 1587 & $-19.47\pm 0.11$ & $-1.00\pm 0.09$ & morphological class{.} \\
         &  5550 & $B_\mathrm{Zw}$ & 15.5                     & 0.03       & Spiral         & 3227 & $-19.62\pm 0.08$ & $-1.11\pm 0.07$ & \\ 
         &  5550 & $B_\mathrm{Zw}$ & 15.5                     & 0.03       & Irr-Pec        &  204 & $-20.05\pm 0.05$ & $-1.81\pm 0.24$ & \\ 
 DARS1   &  70.3 & $b_\mathrm{J}$  & 17.0                     & 0.06       & E-S0           &   97 & $-19.12\pm 0.20$ & $-0.48\pm 0.25$ & morphological class{.} \\
         &  70.3 & $b_\mathrm{J}$  & 17.0                     & 0.06       & Sp-Irr         &  194 & $-19.48\pm 0.20$ & $-1.24\pm 0.25$ & \\ 
 DARS2   &  70.3 & $B$             & 17.0                     & 0.06       & 1/2-red        &  144 & $-19.40\pm 0.17$ & $-0.20\pm 0.26$ & $(U-B)_\mathrm{rest}>0.2$ \\
         &  70.3 & $B$             & 17.0                     & 0.06       & 1/2-blue       &  144 & $-19.55\pm 0.23$ & $-1.55\pm 0.17$ & $(U-B)_\mathrm{rest}<0.2$ \\
 DUKST   &  1460 & $b_\mathrm{J}$  & 17.0                     & 0.07       & ALL            & 2500 & $-19.52\pm 0.10$ & $-1.04\pm 0.08$ & \\        
 SAPM $^{\mathrm{a}}$   &  4300 & $b_\mathrm{J}$  & 17.15     & 0.07       & E-S0           &  311 & $-19.46\pm 0.25$ & $\;\;\; 0.20\pm 0.35$ & morphological class{.} \\
         &  4300 & $b_\mathrm{J}$  & 17.15                    & 0.07       & Sp-Irr         &  999 & $-19.24\pm 0.16$ & $-0.80\pm 0.20$ & \\
         &  4300 & $b_\mathrm{J}$  & 17.15                    & 0.07       & H$\alpha$-low  &  599 & $-19.38\pm 0.24$ & $-0.75\pm 0.28$ & EW(H$\alpha$) $<$ 2 \AA \\ 
         &  4300 & $b_\mathrm{J}$  & 17.15                    & 0.07       & H$\alpha$-mid  &  473 & $-19.09\pm 0.23$ & $-0.72\pm 0.29$ & 2 $<$ EW(H$\alpha$) $<$ 15\AA \\
         &  4300 & $b_\mathrm{J}$  & 17.15                    & 0.07       & H$\alpha$-high &  459 & $-18.93\pm 0.26$ & $-1.28\pm 0.30$ & EW(H$\alpha$) $>$ 15 \AA \\
 Autofib &  10.2 & $b_\mathrm{J}$  & 24.0                     & 0.02--0.15 & Red-E          &  154 & $-20.25\;\;\;\;\;\;\;\;\;\;$ & $-0.92\;\;\;\;\;\;\;\;\;\;$ & spectral class{.} by \\
         &  10.2 & $b_\mathrm{J}$  & 24.0                     & 0.02--0.15 & Blue-E         &  177 & $-19.29\;\;\;\;\;\;\;\;\;\;$ & $-0.94\;\;\;\;\;\;\;\;\;\;$ & cross-correlation \\
         &  10.2 & $b_\mathrm{J}$  & 24.0                     & 0.02--0.15 & Sab            &  282 & $-19.88\;\;\;\;\;\;\;\;\;\;$ & $-1.19\;\;\;\;\;\;\;\;\;\;$ & \\
         &  10.2 & $b_\mathrm{J}$  & 24.0                     & 0.02--0.15 & Sbc            &  361 & $-19.20\;\;\;\;\;\;\;\;\;\;$ & $-1.26\;\;\;\;\;\;\;\;\;\;$ & \\
         &  10.2 & $b_\mathrm{J}$  & 24.0                     & 0.02--0.15 & Scd            &  539 & $-19.04\;\;\;\;\;\;\;\;\;\;$ & $-1.40\;\;\;\;\;\;\;\;\;\;$ & \\
         &  10.2 & $b_\mathrm{J}$  & 24.0                     & 0.02--0.15 & Sdm            &   90 & $-18.82\;\;\;\;\;\;\;\;\;\;$ & $-1.44\;\;\;\;\;\;\;\;\;\;$ & includes Starburst \\
 2dFGRS1 & $\sim60$ & $b_\mathrm{J}$  & 19.45                 & 0.15       & Type-1         & 1850 & $-19.36\pm 0.09$ & $-0.74\pm 0.11$ & PCA-spectral class{.} \\
         & $\sim60$ & $b_\mathrm{J}$  & 19.45                 & 0.15       & Type-2         &  958 & $-19.46\pm 0.14$ & $-0.86\pm 0.15$ & \\
         & $\sim60$ & $b_\mathrm{J}$  & 19.45                 & 0.15       & Type-3         & 1200 & $-19.22\pm 0.12$ & $-0.99\pm 0.13$ & \\
         & $\sim60$ & $b_\mathrm{J}$  & 19.45                 & 0.15       & Type-4         & 1193 & $-18.86\pm 0.12$ & $-1.21\pm 0.12$ & \\
         & $\sim60$ & $b_\mathrm{J}$  & 19.45                 & 0.15       & Type-5         &  668 & $-18.94\pm 0.22$ & $-1.73\pm 0.16$ & \\
 2dFGRS2 & $\sim1200$  & $b_\mathrm{J}$  & 19.45              & 0.15       & Type-1         & 27540& $-19.33\pm 0.05$ & $-0.54\pm 0.02$ & PCA-spectral class{.} \\
         & $\sim1200$  & $b_\mathrm{J}$  & 19.45              & 0.15       & Type-2         & 24256& $-19.37\pm 0.03$ & $-0.99\pm 0.01$ & \\
         & $\sim1200$  & $b_\mathrm{J}$  & 19.45              & 0.15       & Type-3         & 15016& $-19.03\pm 0.04$ & $-1.24\pm 0.02$ & \\
         & $\sim1200$  & $b_\mathrm{J}$  & 19.45              & 0.15       & Type-4         &  8386& $-19.07\pm 0.05$ & $-1.50\pm 0.03$ & \\ 
 ESP     &  23.2 & $b_\mathrm{J}$  & 19.4                     & 0.15       & no-emi         & 1767 & $-19.37\pm 0.10$ & $-0.98\pm 0.09$ & EW $<$ 5 \AA \\
         &  23.2 & $b_\mathrm{J}$  & 19.4                     & 0.15       & emi-line       & 1575 & $-19.31\pm 0.11$ & $-1.40\pm 0.10$ & EW $>$ 5 \AA \\
 SDSS    & $\sim 2000$ & $g^*$ & 17.69                        & 0.02--0.17 & ALL            &53999 & $-18.96\pm 0.02$ & $-0.90\pm 0.03$ & \\
 Norris  &  25.0 & $B_\mathrm{AB}$ & $r\le20.0$               & 0.0--0.2   & no-OII         &  159 & $-19.83\pm 0.46$ & $-1.09\pm 0.23$ & EW[OII] $<$ 10 \AA\\
         &  25.0 & $B_\mathrm{AB}$ & $r\le20.0$               & 0.0--0.2   & OII            &   60 & $-19.12\pm 1.08$ & $-1.10\pm 0.45$ & EW[OII] $>$ 10 \AA\\
COMBO17$^{\mathrm{b}}$  &  0.78  & $B$ & $R \la24.0$ & 0.2--0.4   & Type-1        &  344 & $-19.19\pm 0.15$ & $\;\;\;0.62\pm0.20$ & fits of obs. SEDs \\
         &  0.78  & $B$ & $R\la24.0$ & 0.2--0.4   & Type-2        &  986 & $-19.42\pm 0.17$ & $-0.80\pm 0.08$ & of redshifted temp.\\
         &  0.78  & $B$ & $R\la24.0$ & 0.2--0.4   & Type-3        & 1398 & $-19.81\pm 0.23$ & $-1.24\pm 0.07$ & \\
         &  0.78  & $B$ & $R\la24.0$ & 0.2--0.4   & Type-4        & 2946 & $-18.75\pm 0.16$ & $-1.37\pm 0.06$ & \\
 CFRS    & 0.0347& $B_\mathrm{AB}$ & $I_\mathrm{AB}<22.5$     & 0.2--0.5   & 1/2-red        &   -  & $-18.79\pm 0.35$ & $\;\;\; 0.00\pm 0.20$ & $(V$-$I)_\mathrm{AB}$ of redshifted \\
         & 0.0347& $B_\mathrm{AB}$ & $I_\mathrm{AB}<22.5$     & 0.2--0.5   & 1/2-blue       &   -  & $-19.18\pm 0.35$ & $-1.34\pm 0.20$ & non-evolv{.} Sbc temp{.} \\
 CADIS   & 0.0833& $B$             & $I_{815}\la 23$          & 0.3--0.5   & E-Sa           &   82 & $-19.29\pm 0.22$ & $\;\;\; 0.19\pm 0.22$ & fit of obs{.} SEDs \\
         & 0.0833& $B$             & $I_{815}\la 23$          & 0.3--0.5   & Sa-Sc          &  301 & $-19.66\pm 0.30$ & $-0.81\pm 0.13$ & to redshifted temp{.}\\
         & 0.0833& $B$             & $I_{815}\la 23$          & 0.3--0.5   & Starburst      &  252 & $-18.69\pm 0.29$ & $-1.28\pm 0.21$ & \\
 CNOC2   & 0.692 & $B_\mathrm{AB}$ & $R_\mathrm{c} < 21.5$    & 0.55       & Early          &  611 & $-18.92\pm 0.12$ & $\;\;\; 0.08\pm 0.14$ & least-square fit of obs{.} \\
         & 0.692 & $B_\mathrm{AB}$ & $R_\mathrm{c} < 21.5$    & 0.55       & Interm         &  518 & $-19.24\pm 0.16$ & $-0.53\pm 0.15$ & $UB_\mathrm{AB}VR_CI_C$ colors \\                 
         & 0.692 & $B_\mathrm{AB}$ & $R_\mathrm{c} < 21.5$    & 0.55       & Late           & 1016 & $-19.12\pm 0.16$ & $-1.23\pm 0.12$ & to redshifted temp{.} \\             
 CNOC1   &     - & $B_\mathrm{AB}$ & $r\le22.0$               & 0.2--0.6   & 1/2-red        &  209 & $-19.32\pm 0.30$ & $-0.38\pm 0.29$ & $r$-$g$ of redshifted \\
         &     - & $B_\mathrm{AB}$ & $r\le22.0$               & 0.2--0.6   & 1/2-blue       &  180 & $-19.71\pm 0.50$ & $-1.44\pm 0.32$ & non-evolv{.} Sbc temp{.}\\
 ESS     & 0.219 & $B$             & 22.0                     & 0.1--0.6   & Early          &  108 & $-19.52\pm 0.24$ & $-0.24\pm 0.33$ & PCA-spectral class{.} \\
         & 0.219 & $B$             & 22.0                     & 0.1--0.6   & Interm         &  154 & $-19.37\pm 0.20$ & $-0.75\pm 0.21$ & \\
         & 0.219 & $B$             & 22.0                     & 0.1--0.6   & Late           &  190 & $-19.00\pm 0.20$ & $-1.25\pm 0.20$ & \\
\hline
\end{tabular}
\smallskip
\\
\end{center}
\end{table*}

\begin{table*}
\begin{list}{}{}
\item[\underbar{Table notes}:]
\item[-] See Table \ref{mstar_alpha_UV_tab} for definition of solumns. All references are provided in the text.
\item[-] All listed values of $M^*$ result from the conversion from the original values derived by 
         the authors in the filters listed in column (3), into the Johnson $B$ band. 
         The original values of $\alpha$ are kept unchanged.
\item[-] All quoted $b_\mathrm{J}$ magnitudes are from photographic plates.
\item[$^{\mathrm{a}}$] In the survey denoted SAPM, sub-sample
H$\alpha$-low contains 233 E-S0, 217 Sp-Irr, 149 unclassified
galaxies; sub-sample H$\alpha$-mid, 24 E-S0, 358 Sp-Irr, and 81
unclassified galaxies; sub-sample H$\alpha$-high, 20 E-S0, 344 Sp-Irr,
and 95 unclassified galaxies.
\item[$^{\mathrm{b}}$] The values of $M^*$ and $\alpha$ for the
COMBO-17 survey are converted from a cosmology with [$\Omega_m=0.3$,
$\Omega_\lambda=0.7$] into [$\Omega_m=1.0$, $\Omega_\lambda=0.0$]
using empirical corrections described in the text. These values should
therefore be used with caution.
\end{list}
\end{table*}

The most numerous measurements of intrinsic LFs were obtained in the
$B$ band. For clarity, \fg \ref{mstar_alpha_B} shows the $M^*$ and
$\alpha$ parameters for samples grouped in four intervals of effective
depth $z_\mathrm{max}$. 

\subsubsection{$B$ luminosity functions at redshifts 0.4--0.6          \label{Bcomp_high}}

The lower-left panel of \fg \ref{mstar_alpha_B} shows the intrinsic LF
parameters for the redshift surveys providing measurements at
$z_\mathrm{max}\sim 0.6$: the Canada-France Redshift Survey
\citep[denoted CFRS]{lilly95}; the CNOC1 \citep{lin97}, CNOC2
\citep{lin99}, COMBO-17 \citep{wolf03}, and ESS \citep{lapparent03a},
already mentioned in \scts\ref{Ucomp}, \ref{Vcomp} and \ref{Rcomp};
and the Calar Alto Deep Imaging Survey \citep[denoted
CADIS]{fried01}. The photometric catalogues on which are based all
these redshift surveys are obtained from CDD imaging (see Table
\ref{mstar_alpha_B_tab} for the sample parameters).

As in the $U$ and $R_\mathrm{c}$ bands, I use the listed values of
$M^*(z=0.3)$ for the CNOC2 $B$ LFs \citep{lin99}, and the intrinsic
LFs in the interval $0.2 \la z \la 0.4$ for the COMBO-17 survey
\citep{wolf03}. As in the $V$ and $R_\mathrm{c}$ bands, I use for the
ESS the values of $M^*$ and $\alpha$ derived from the full redshift
range of the survey ($0.1\le z\le0.6$).  Both the CADIS and CFRS
measure LFs in the intervals $0.5 \la z \la 0.75$ and $0.75 \la z \la
1.0$, in which they detect evolutionary effects using the LFs with
$z\le0.5$ as reference. I however use their intrinsic LFs derived in
the intervals $0.3\le z\le0.5$ and $0.2\le z\le0.5$ \respn, because
these provide the best constraint on $\alpha$ for each survey
\citep[see][]{fried01,lilly95}; these redshift intervals also 
correspond to $z_\mathrm{max}\la0.6$.  For the CFRS, the values of
$M^*$, listed for $h=0.5$ \citep{lilly95}, are converted to $h=1$.
Note that no uncertainties are quoted by \citet{lilly95} for the
Schechter parameters of the CFRS 1/2-red and 1/2-blue LFs. In Table
\ref{mstar_alpha_B}, I have approximated these uncertainties as
$\sqrt{2}$ times the uncertainties $\sigma(M)=0.25$ and
$\sigma(\alpha)=0.15$ quoted for the general LF in the interval
$0.2\le z\le0.5$ \citep[see \sct 3.1.1 in][]{lilly95}. The values of
$M^*$ for the CNOC2, CNOC1, and CFRS are measured in $B_\mathrm{AB}$:
I convert them into Johnson $B$ magnitudes using
$B-B_\mathrm{AB}=0.14$, as estimated by \citet{fukugita95}.

I first compare the LFs for the 3 surveys which are based on a
spectral classification, and are split into 3 spectral classes: the
ESS, CNOC2 and CADIS.  Despite selection effects specific to each
sample, \fg \ref{mstar_alpha_B} shows that the $B$ band intrinsic LFs
for the 3 surveys have a similar behavior. All 3 surveys show the
steepening in $\alpha$ from early to late-type classes: the slope
$\alpha$ increases from values inside the interval
$-0.24\le\alpha\le0.19$ for the early-type galaxies, to the narrow
range $-1.28\le\alpha\le-1.23$ for the late-type galaxies.  The values
of $M^*$ also describe narrow intervals of $\la0.6 ^\mathrm{mag}$
among the 3 surveys for each of the 3 classes.  If one assumes that
the 3 spectral classes in the ESS, CNOC2 and CADIS sample similar
galaxy populations, taken 2 by 2, the ESS and CADIS LFs, and the CNOC2
and CADIS LFs are in agreement at the 1-$\sigma$ level.

There are however 2 noticeable differences between the CNOC2 and ESS
$B$ LFs. First, there is a 2-$\sigma$ difference between the $M^*$
values for the CNOC2 and ESS early-type LFs.  I also note that there
is only a very small dimming of the bright-end of the CNOC2 LF from
early to late types: a shift of approximately $0.5 ^\mathrm{mag}$
towards faint magnitudes is however evaluated from \fg 5 of
\citet{lin99}; it converts into a $0.2 ^\mathrm{mag}$ brightening of
$M^*(B)$ from early to late types because of the correlation between
$M^*$ and $\alpha$ (see \sct \ref{Vcomp}). In contrast, there is a
$\simeq1.0 ^\mathrm{mag}$ shift towards faint magnitudes of the
bright-end of both the ESS and CADIS $B$ LFs from early to late types,
which is measured by a dimming in $M^*(B)$ of $0.52 ^\mathrm{mag}$ in the ESS and
$0.60 ^\mathrm{mag}$ in the CADIS.  These effects could be due
to the incompleteness of the ESS $B$ sample, and a difference in
morphological type content of the spectral classes in the 3 surveys.

Because the spectral classification and redshift measurement in the
CADIS and COMBO-17 surveys are based on a similar technique
(medium-band photometry; see \sct \ref{Ucomp} and below), it is
useful to compare the results from the 2 surveys. The major
differences between the 2 surveys are the larger statistic for the
COMBO-17 and its use of 4 spectral classes, compared to 3 spectral
classes in the CADIS (see Table \ref{mstar_alpha_B_tab}; there is also
some difference in the set of filters, see below).  Lower-left panel
of \fg \ref{mstar_alpha_B} shows that there is remarkable agreement
between the CADIS E-Sa, Sa-Sc, Starburst LFs and the COMBO-17 Type-1
(E-Sa), Type-2 (Sa-Sbc), Type-4 (SB6-SB1) LFs \respn, as expected from
the similar spectral type content in the corresponding classes.  As
observed in the ESS and CADIS $B$ LFs, the LF for COMBO-17 Type-4
galaxies shows the dimming of $M^*$ due to the expected dominant
contribution from dI galaxies in this class.  The noticeable
brightening in $M^*$ for the COMBO-17 $B$ LF for Type-3 (Sbc-SB6)
galaxies, compared to the Type-1 and Type-2 classes, may be due to the
combination of significant mass and star formation rate in $B$ for the
galaxies in this class.  Here, contrary to the $U$ and $R_\mathrm{c}$
bands, there is no need to invoke some systematic bias in the absolute
magnitudes (related to the larger redshifts errors for the faint
Starburst galaxies). Note that the Johnson $B$ band is the only band
considered here in which the COMBO-17 LFs are directly provided.  This
strengthens the suggestion that the difference in the $U$ and
$R_\mathrm{c}$ bands between the COMBO-17 Type-2 and Type-3 LFs and
the intermediate-class LFs of the CNOC2 and ESS may be due to some
biases in the color coefficients required to convert the COMBO-17
$m_{280}$ and $r^*$ LFs into the $U$ and $R_\mathrm{c}$ bands
respectively.

The type content and respective fractions of galaxies in the CADIS
classes (13\% E-S0, 47\% Sa-Sc, 40\% Starburst galaxies) and the
COMBO-17 classes (6\% Type-1 [E-Sa], 17\% Type-2 [Sa-Sbc], 25\% Type-3
[Sbc-SB6], 52\% Type-4 [SB6-SB1] galaxies) also indicate that the
CADIS class Sa-Sc is expected to contain a significant part of both
the COMBO-17 Type-2 and Type-3 galaxies. Taking the average of the
Schechter parameters for the COMBO-17 Type-2 and Type-3 classes yields
values of $M^*$ and $\alpha$ in acceptable agreement with the CADIS
Sa-Sc class (at less than 1-$\sigma$ level for $M^*$, and less than
2-$\sigma$ level for $\alpha$). The $B$ LFs for the COMBO-17 therefore
demonstrate the gain in information when changing from 3 to 4 spectral
classes. This gain is effective because the chosen COMBO-17 classes
succeed in separating galaxies with different intrinsic LFs.

The agreement of the CADIS and COMBO-17 $B$ LFs with those for the ESS
and CNOC2 demonstrates the interest of the ``photometric redshift''
approach for measuring LFs: the CADIS survey is based on a combination
of 4 wide-band filters ($BRJK^\prime$) and up to 13 medium-band
($\Delta\lambda\simeq250-500$\AA) filters; the redshifts and spectral
types of $\sim 2780$ galaxies were measured using a standard
minimization procedure in which the observed SEDs are compared to a
spectral library. The resulting redshifts uncertainties are
$\sigma(z)\le0.03$, to be compared to $\sim0.0001-0.003$ for the
spectroscopic surveys listed in Tables
\ref{mstar_alpha_UV_tab}--\ref{mstar_alpha_R_tab} (for comparison, the
COMBO-17 uses 5 wide-band filters --Johnson $UBVRI$-- and 12
medium-band filters with FWHM $\simeq140-310$\AA\ which yield the same
redshift uncertainty as in the CADIS).  The $\sigma(z)\le0.03$
redshift uncertainty in the CADIS survey
is nevertheless sufficient for derivation of spectral-type LFs in
agreement with those derived from redshift surveys such as the CNOC2
and ESS. The reason is that the dispersion in the absolute magnitudes
caused by the redshift uncertainties, of order of $5
\sigma(z)/z$, are significantly smaller than the width of the LF for
each morphological types (see \fg \ref{LFlocal_fig} in \sct
\ref{LFlocal}): for example, $5 \sigma(z)/z = 0.3$ at $z=0.5$ in the
CADIS survey, which is nearly 10 times smaller than the dispersion of
the Gaussian LFs for the giant galaxies (see Table \ref{LFlocal_tab}
in \sct \ref{LFlocal}); this is even smaller than the 8 magnitude
interval over which the Schechter LFs for dwarf galaxies are defined
\citep[see][]{jerjen00,trentham02b}.

The 2-class LFs derived from the CNOC1 and CFRS, also plotted in the
lower-left panel of \fg \ref{mstar_alpha_B}, are based on a color cut
at the redshifted color of a non-evolving Sbc galaxy. Both samples
confirm the steepening in $\alpha$ for bluer galaxies observed for the
ESS, CNOC2, CADIS and COMBO-17 surveys. It is however noticeable that
for both the CFRS and the CNOC1, $M^*$ for the blue sample is $\sim0.4
^\mathrm{mag}$ brighter than for the red sample. Examination of the
corresponding curves in \citet{lin97} and \cite{lilly95} shows that
this effect is due to the correlation between $M^*$ and $\alpha$ in
the Schechter parameterization: for the CNOC1, the bright-end of the
blue LF is actually \emph{fainter} by $\sim0.2-0.3 ^\mathrm{mag}$ than
that for the red LF; for the CFRS, the bright-end of the blue LF is
not determined, but the few common points with the red LF (those lying
around the ``knee'' at $M(B_\mathrm{AB})\simeq-19.5$) suggest also a
\emph{fainter} bright-end by $\la 0.5 ^\mathrm{mag}$ for the blue
LF. This confirms the $\alpha$-dependent relation mentioned in \sct
\ref{Vcomp} between the bright exponential fall-off of a given LF and
the value of $M^*$.

The $0.2$ to $0.5 ^\mathrm{mag}$ dimming of the LF bright-end from the
CNOC1 and CFRS 1/2-red to 1/2-blue samples is however smaller than for
the ESS ($\sim1.0 ^\mathrm{mag}$). The small dimming in $M^*$ for the
CNOC1 and CFRS in the $B$ band is similar to that already described
for the CNOC1 LFs in the $r$ band in \sct \ref{Rcomp} and attributed
to type-mixing: the use of only 2 spectral classes fails in separating
the blue low luminosity galaxies from the more luminous Spiral
galaxies; the bright ends of the red and blue LFs are dominated by
Elliptical and Spiral galaxies \respn, which have similar
characteristic magnitudes (see Table \ref{LFlocal_tab} and \fg
\ref{delta_MVR}); due to the correlation between $M^*$ and $\alpha$,
combination with a steeper $\alpha$ for the blue LF then results in a
brighter $M^*$ for that LF.  As in the $R_\mathrm{c}$ and $V$ bands,
comparison of the ESS, CNOC2 and CADIS LFs with those for the CNOC1
and CFRS illustrates the significant gain of information when changing
from 2 to 3 spectral classes, due to the multiplicity of shape for the
intrinsic LFs of the dominant morphological types (see \sct
\ref{LFlocal}).

\subsubsection{$B$ luminosity functions at redshifts 0.15--0.20        \label{Bcomp_mid_high}}

The lower-right panel of \fg \ref{mstar_alpha_B} shows the $M^*$ and
$\alpha$ parameters for the intrinsic LFs measured from redshift
surveys with $0.15\la z_\mathrm{max}\la 0.20$: the first and second
sub-samples of the 2dF Galaxy Redshift Survey (2dFGRS) from which
intrinsic LFs were derived, based on 5869 galaxies \citep[denoted
2dFGRS1]{folkes99}, and 75589 galaxies \citep[denoted
2dFGRS2]{madgwick02a}; the ESO Slice Project \citep[denoted
ESP]{zucca97}; the Norris Survey of the Corona Borealis Supercluster
\citep{small97b}. The photometric surveys on which are based all these
mentioned redshift surveys are obtained from digitized photographic
plates.

For converting the values of $M^*$ measured as $b_\mathrm{J}$
magnitudes into Johnson $B$ magnitudes for the 2dFGRS1, 2dFGRS2, and
ESP, I apply the $B-b_\mathrm{J}=0.28(B-V)$ color equation determined
by \citet[][see also \citealt{norberg02}]{blair82} for the UK Schmidt
Telescope photographic system, complemented by the $B-V$ colors
estimated by \citet[Table 3a]{fukugita95}: for the 5 2dFGRS1 types
listed in Table \ref{mstar_alpha_B_tab}, I use the average $B-V$ color
$0.905$ over listed types E and S0, and the $B-V$ colors 0.78, 0.57,
0.50, 0.27, for listed types Sab, Sbc, Scd, Im \respn; for the 4
2dFGRS2 types, I use the average $B-V$ color $0.905$ over listed types
E, S0 and Sa, and the $B-V$ colors 0.57, 0.50, and 0.27, for listed
types Sbc, Scd, Im \respn; for the ESP, I use the average $B-V$ color
$0.905$ over listed types E and S0, and 0.57 for listed type Sbc. The
resulting $B-b_\mathrm{J}$ colors are 0.25 for the average between
listed types E and S0, and 0.22, 0.16, 0.14, 0.08 for listed types
Sab, Sbc, Scd, Im \respn, which are assigned to the 2dFGRS1; 0.25 for
the average between listed types E and S0, and 0.16, 0.14, 0.08 for
listed types Sbc, Scd, Im \respn, which are assigned to the 2dFGRS2;
0.25 for the average between listed types E and S0, and 0.16 for
listed type Sbc, which are assigned to the ESP.

For the Norris survey, I convert values of $M^*$ measured in
$B_\mathrm{AB}$ into Johnson $B$ magnitudes using again
$B-B_\mathrm{AB}=0.14$, as estimated by \citet{fukugita95}.  Note that
the areas of sky sampled by the 2dFGRS1 and 2dFGRS2 are not provided
by the authors; I roughly estimate them using the other elements of
information provided by the authors (number of fields and number of
spectra per field). The resulting approximate areas listed in Table
\ref{mstar_alpha_B_tab} could be in error by as much as a factor 2.

In the ESP \citep{zucca97}, the detection/no-detection of the
[OII]$\lambda$3727 emission line is used for separating the sample
into 2 spectral classes; as stated by the authors and indicated in
Table \ref{mstar_alpha_B_tab}, detection of emission lines corresponds
to a threshold of 5\AA\ in equivalent width. The resulting LF for the
galaxies with no or weak [OII] line has a nearly flat slope in the $B$
band ($\alpha\simeq-1.0$; the corresponding point in \fg
\ref{mstar_alpha_B} is overlayed with that for the 2dFGRS2 Type-2 LF),
and a steeper slope is measured for the galaxies with strong [OII]
($\alpha\simeq-1.3$); the variation in $M^*$ is small from one
sub-sample to the other and within the error bars. As for the LCRS $r$
LFs \citep{lin96} based on the equivalent width of the [OII] line (see
\sct \ref{Rcomp}), I interpret the flat slope of the ESP LF for
galaxies with low [OII]-emission as the result of type mixing, as
demonstrated in \fg \ref{delta_WOII} (see \sct \ref{Rcomp_low}): the
LF for that sample is likely to be contaminated at the faint end by
Spiral and Irregular galaxies, thus failing to isolate the bounded LFs
for E and S0 galaxies. Note however that both ESP LFs in the $B$ band
have a steeper value of $\alpha$ by $\sim0.5$ compared to the
corresponding LCRS LFs in the $r$ band. This effect might be due to
the bias against low surface brightness galaxies which affects the
LCRS, and tends to exclude late-type galaxies. As a result, the low
and high [OII]-emission galaxy classes in the LCRS may be shifted
towards earlier types. Comparison of LFs among different filters must
however be taken with caution.

In the Norris survey \citep{small97b}, shown in the lower right panel
of \fg \ref{mstar_alpha_B}, the 2 intrinsic LFs are also estimated
using the strength of the [OII]$\lambda$3727 emission line. Although a
$\sim 0.7 ^\mathrm{mag}$ dimming of $M^*$ is observed for galaxies with
EW[OII] $>$ 10 \AA~ compared to those with EW[OII] $>$ 10 \AA, no
change in the slope $\alpha$ is observed between the 2 sub-samples,
probably due to poor statistics (see the large error bars). Although
the Norris survey reaches $z_\mathrm{max}\simeq0.5$, here I only
consider the LFs for the sub-samples with $0< z\le0.2$, as the slope
$\alpha$ for the $0.2< z\le0.5$ sub-samples is poorly determined (they
only include galaxies brighter than $M(B)\la -19$). The Norris LFs
could be improved by extending the [OII]-line sub-samples to the full
redshift range $0<z\le0.5$, thus doubling the number of galaxies per
sub-sample \citep[see][]{small97b}.

In contrast to the ESP and Norris surveys, the spectral
classifications for the 2dFGRS1 and 2dFGRS2, whose LFs are also shown
in the lower-right panel of \fg \ref{mstar_alpha_B}, are based on a
PCA, and use the projections onto the first 2 principal components
(after exclusion of the mean spectrum).  The 2dFGRS1 is separated into
5 types, which I estimate to correspond to morphological types E/S0,
Sab, Sbc, Scd, and Sdm/Im \resp \citep[see \fg 8 of][]{folkes99}.  The
2dFGRS2 is divided into 4 types; from \fg 4 of \citet{madgwick02a}, I
estimate that they correspond to morphological types E/S0/Sa,
Sa/Sb/Scd, Sb/Scd, and Scd/Sm/Im respectively.\footnote{Although
\citet{kennicutt92} galaxies with types later than Scd are not
represented by \citealt{madgwick02a}, I assume that these objects
would be included in the latest class.}

The usual systematic steepening of the intrinsic LFs for later type
galaxies is observed in both the 2dFGRS1 and 2dFGRS2 samples, and the
2 samples describe consistent intervals of $\alpha$ (within the error
bars). Both samples also show a dimming in $M^*$ between the LFs for
the earliest and the latest class (by $\sim0.6 ^\mathrm{mag}$ for the
2dFGRS1, and $\sim0.4 ^\mathrm{mag}$ for the 2dFGRS2). Nonetheless,
both the 2dFGRS1 and 2dFGRS2 fail to detect the Gaussian shape of the
intrinsic LFs for E, S0 and Sa galaxies.  Because the 2dFGRS1 and
2dFGRS2 samples reach absolute magnitudes as faint as $M(B)\simeq-16$
and $M(B)\simeq-14$ \respn, both surveys should a priori detect the
fall-off for the E, S0 and Spiral intrinsic LFs at faint magnitude
(\citealp{jerjen97b}; see also \fg \ref{LFlocal_fig} above).  A dip at
$-17.5\la M(b_\mathrm{J})\la-16.5$ is actually visible in the Type-1
LF for the 2dFGRS1 \citep[see \fg 11 of][]{folkes99}, and calls for
confirmation with a larger sample.  Although the 2dFGRS2 sample is
$\sim 20$ times larger than the 2dFGRS1, the 2dFGRS2 Type-1 LF only
shows a weak minimum at $M(b_\mathrm{J})\sim-16.0$. The different
behavior between the 2 samples is probably due to the different
definition of the spectral types.

Note that the 2dFGRS1 and 2dFGRS2 samples are separated into 1 and 2
more spectral classes \resp than the ESS and CNOC2; as a result, one
would expect that their respective Type-1 samples show an even lower
degree of morphological type mixing than in the ESS and CNOC2.  The
situation may however be opposite. As mentioned above for the LCRS,
multi-fiber spectroscopy results in systematic color biases due to the
small circular apertures, and in large random errors due to the
inaccurate flux-calibration of the fibers and their inaccurate
positioning onto the objects. Moreover, the design of the corrector
lens of the 2dF multi-fiber spectrograph causes a chromatic
displacement of different components of a given spectrum
\citep{madgwick02a}. The PCA of the 2dFGRS1 uses the flux-calibrated
spectra using an average response curve of the instrument; this curve
shows wavelength-dependent variations as large as $\sim 20$\%
\citep{folkes99}, and does not account for the fiber-to-fiber and time
variations, which cause additional dispersion in the flux-calibration.
The 2dFGRS1 Type-1 classes may therefore be contaminated by galaxies
with later spectral-types.  As these have a nearly flat faint-end
slope, the contamination tends to erase the Gaussian behavior of the E
and S0 included in this class. In the 2dFGRS2, the Type-1 class
contains predominantly E, S0 and Sa galaxies \citep[see \fg 4
of][]{madgwick02a}, which also have Gaussian LFs (see \sct
\ref{LFlocal}). The incompatibility of the 2dFGRS2 Type-1 LF
\citep[see \fg 11 of][]{madgwick02a} with a Gaussian LF suggests that
this sample is also affected by contamination among the spectral
types.

The intrinsic LFs estimated from the commissioning data of the SDSS
and based on rest-frame colors \citep{blanton01} do not appear to be
affected by these effects. Because \citet{blanton01} do not provide
the Schechter parameters fitted to these LFs, they are not plotted in
\fg \ref{mstar_alpha_B_tab}. From visual inspection of \fg 14 of
\citet{blanton01}, the LFs in the $r^*$ band for the rest-frame color
intervals $0.74< g^*-r^* <0.90$ and $0.58< g^*-r^* <0.74$, which
correspond to morphological types E, and S0/Sa galaxies \resp
\citep[see][]{fukugita95} show a clear fall-off at faint magnitudes,
in the intervals $-20.0\la M(r^*)\la -18.3$, and $-19.0\la M(r^*)\la
-16.7$ respectively. I suggest that the SDSS LFs are able to detect
the bounded behavior at faint magnitude for the giant galaxies (E, S0,
and Spiral) because these LFs are based on rest-frame colors and
\emph{not} on spectral classification. As the SDSS also uses
multi-fiber spectroscopy, a spectral classification based on these
data would likely be affected by aperture bias and calibration errors.

The 2dFGRS1 and 2dFGRS2 surveys show the same effect as observed for
the ESS intermediate-type LF: nearly flat slopes are measured for the
2dFGRS1 Type-2 LF ($\alpha=-0.86$; this spectral class corresponds to
morphological type Sab) and Type-3 LF ($\alpha=-0.99$; corresponding
to type Sbc), and for the 2dFGRS2 Type-2 LF ($\alpha=-0.99$; corresponding
to types Sa/Sb/Scd). In the ESS, these flat slopes are reconciled with the
Gaussian shapes of the intrinsic LF for Spiral galaxies by adding a
contribution from dwarf spheroidal galaxies \citep[see \fg 11
in][]{lapparent03a}, which is justified by the bluer colors of the
dSph galaxies as compared to their giant analogs (E and S0 type). I
propose a similar interpretation of the flat slopes of the
intermediate-type LFs for the 2dFGRS1 and 2dFGRS2 surveys; it could
also apply to the SDSS LF for rest-frame colors $0.42< g^*-r^* <0.58$
(corresponding to Sbc/Scd galaxies; see \citealp{fukugita95}), which
has a flat faint-end slope \citep[see \fg 14 in][]{blanton01}. This
interpretation is at variance with that of \citet{kochanek01b}, who
show that in redshift surveys based on multi-fiber spectroscopy, the
mix of the various morphological classes yields a false artificial
steep slope for the Spiral galaxies: the interpretation of
\citet{kochanek01b} ignores the Gaussian behavior of the Spiral
intrinsic LF.

Note that the SDSS LF for the earliest class ($0.74< g^*-r^* <0.90$)
shows an upturn at $M(r^*)\ga-18.0$ \citep{blanton01}. I already
mentioned that the Type-1 LFs for the 2dFGRS1 and 2dFGRS2 show an
upturn at $M(b_\mathrm{J})\ga-17.0$ and $M(b_\mathrm{J})\ga-16.0$
respectively. These upturns could be explained by a population of red
dSph galaxies, as detected in the Coma cluster
\citep[see][]{andreon01c}. However, at these faint magnitudes, the
signal in the early-type LFs for the SDSS and 2dF samples is of low
significance, and calls for caution in its interpretation; if such a
population exists, it appears of lower density than the population of
bluer dSph galaxies which presumably flattens the intermediate-type
LFs in the ESS \citep{lapparent03a} and could also play the same role
in the SDSS, 2dFGRS1 and 2dFGRS2.

The slopes $\alpha=-1.21$ and $\alpha=-1.24$ for the Type-4 LF in the
2dFGRS1 (corresponding to morphological types Scd) and the Type-3 LF
in the 2dFGRS2 (corresponding to types Sb/Scd) \respn, are also both
symptomatic of type mixing, as galaxies with these morphological types
are expected to have Gaussian LFs \citep[see][and \fg
\ref{LFlocal_fig}]{jerjen97b}.  Therefore, it is likely that these
spectral classes are contaminated by the later type galaxies
(Sm/Im). As shown for the ESS late-type LF \citep{lapparent03a},
combination of a Gaussian LF for the giant galaxies and a steep
Schechter LF for the dwarf galaxies yields a Schechter LF with an
intermediate faint-end slope.  Indeed, steep faint-end slopes
$\alpha=-1.73$ and $\alpha=-1.50$ are measured for the Type-5 LF in
the 2dFGRS1 (corresponding to morphological types Sdm/Im), and the
Type-4 LF in the 2dFGRS2 (corresponding to Scd/Sm/Im)
respectively. The former is in good agreement with the values obtained
for the same types in the nearby redshift surveys CfA2S and SSRS2,
based on morphological classification (see the upper-left panel of \fg
\ref{mstar_alpha_B}, and Table \ref{mstar_alpha_B_tab}; these surveys
are described in \sct\ref{Bcomp_low} below); the flatter slope for the
Type-4 LF in the 2dFGRS2 may again be symptomatic of type mixing.

\subsubsection{$B$ luminosity functions at redshifts 0.02--0.15        \label{Bcomp_mid_low}}

The upper-right panel of \fg \ref{mstar_alpha_B} shows the $M^*$ and
$\alpha$ parameters for the intrinsic LFs measured from redshift
surveys with $0.02\la z_\mathrm{max}\la0.15$: the Stromlo-APM survey
\citep[denoted SAPM]{loveday92,loveday99}, the original
Durham-Anglo-Australian-Telescope Redshift Survey \citep[denoted
DARS1]{efstathiou88a}, and its improved multi-color measurements
\citep[ denoted DARS2]{metcalfe98}; and the Autofib survey
\citep{heyl97}.  The photometric catalogues on which are based all the
mentioned redshift surveys are obtained by digitization of
photographic plates from the UK Schmidt telescope
\citep{efstathiou88a,collins89,maddox90a}.

Again, I convert the values of $M^*$ measured as $b_\mathrm{J}$
magnitudes into the Johnson $B$ band using the
$B-b_\mathrm{J}=0.28(B-V)$ color equation determined by \citet[][see
also \citealt{norberg02}]{blair82}, complemented by the $B-V$ colors
estimated by \citet[Table 3a]{fukugita95}: for the DARS1 and the
2-class SAPM \citep{loveday92}, I use the average $B-V$ color $0.905$
over listed types E and S0, and 0.57 for listed type Sbc; for the
types based on the EW(H$\alpha$) in the SAPM \citep{loveday99}, I use
the average $B-V$ color $0.905$ over listed types E and S0, the
average $B-V$ color $0.675$ over listed types Sab and Sbc, and the
average $B-V$ color $0.385$ over listed types Scd and Im; for the 6
Autofib types, I use the $B-V$ colors 0.96, 0.85, 0.78, 0.57, 0.50,
0.27, for listed types E, S0, Sab, Sbc, Scd, Im respectively. The
resulting $B-b_\mathrm{J}$ colors are 0.25 for the average between
listed types E and S0, and 0.16 for listed type Sbc, assigned to the
DARS1 and the 2-class SAPM \citep{loveday92}; 0.25 for the average
between listed types E and S0, 0.19 for the average between listed
types Sab and Sbc, and 0.11 for the average between listed types Scd
and Im, which are assigned to the SAPM classes based on the
EW(H$\alpha$) \citep{loveday99}; 0.27, 0.24, 0.22, 0.16, 0.14, 0.08
for listed types E, S0, Sab, Sbc, Scd, Im \respn, which are assigned
to the Autofib types.

Although the Autofib survey probes the galaxy distribution to
$z\sim0.75$, the most reliable constraints on the faint-end slope
$\alpha$ of the intrinsic LFs are obtained for $0.02< z\le0.15$
\citep[see \fgs 15, 17 and 20 of][]{heyl97}; the LFs in the intervals
$0.15< z\le0.35$ and $0.35< z\le0.75$ are used by the authors to
constrain the evolution in each LF. Here, I thus use the Schechter
parameters calculated at $z=0.1$ from the parameters given in Table 2
of \citet{heyl97}; the resulting values are listed in Table
\ref{mstar_alpha_B_tab} (note that no uncertainties in the Autofib LFs
are provided by \citealt{heyl97}).

I first describe the intrinsic LFs measured by the Autofib survey
\citep{heyl97}. These LFs describe an even narrower interval of
faint-end slope than the 2dFGRS1 and 2dFGRS2:
$-1.36\le\alpha\le-0.99$. The flat slope ($\alpha\sim-1.0$) measured
for the 3 classes red-E, blue-E, and Sab galaxies means that the
Autofib survey fails to detect the bounded behavior of the early-type
LF at faint magnitudes, which is characterized by $-0.4 \la \alpha \la
0.2$ in the other surveys of the graph. This may also be the result of
a contamination among the galaxy classes. The Autofib survey uses a
spectral classification method based on cross-correlation with a set
of templates. The cross-correlation technique is efficient for
measuring redshifts of absorption-line spectra, as the signal which
builds the cross-correlation peak in a given spectrum is contributed
to by all the absorption lines in that spectrum \citep{tonry79}. In
this approach, the continuum must be subtracted and low-pass filtered,
which is at marked variance with the fact that in a spectral
classification, the dominant signal originates from the \emph{shape}
of the continuum of the spectra \citep{galaz98}. Although there is a
correlation between the absorption line pattern and the continuum of a
spectrum, the absorption lines are sensitive to signal-to-noise, to
the efficiency of the sky subtraction, and to the contamination by OH
sky emission and cosmic rays. When used as a spectral classification,
the cross-correlation technique therefore implies some dispersion due
to the various mentioned effects.

\begin{figure}
  \resizebox{\hsize}{!}
    {\includegraphics{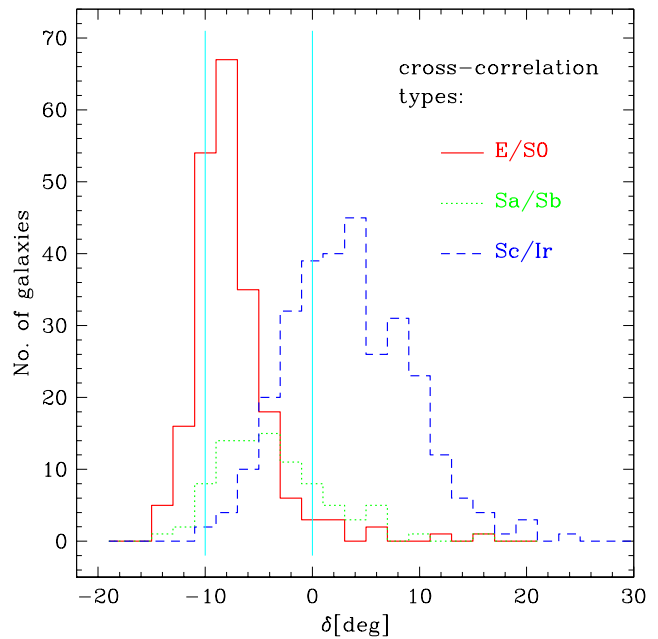}}
\caption{Distribution of the ESS ``cross-correlation'' types as a
function of spectral type $\delta$. The 3 ``cross-correlation''
classes are obtained by cross-correlating with average spectra of
galaxies with morphological types E, S0, Sa, Sb, Sc, and Ir
\citep[from][]{kennicutt92}, and subsequent grouping of the
cross-correlation types in the 3 classes defined by E/S0, Sa/Sb,
Sc/Ir.  The vertical lines mark the corresponding boundaries in
$\delta$ for these 3 morphological classes
\protect\citep{lapparent03a}.}
\label{cross_delta}
\end{figure} 

Here I provide direct evidence that the classification based on
spectrum cross-correlation is responsible for type mixing among the
various classes: the redshifts for the ESS absorption-line spectra
were actually measured by cross-correlation with average
\citet{kennicutt92} template spectra (\citealt{bellanger95a}; see
details on templates in section 2.2 of \citealt{lapparent03a}), thus
providing as a byproduct the cross-correlation types.  I am then able
to compare the ESS cross-correlations types with the PCA spectral
types obtained for the same galaxies. Figure \ref{cross_delta} shows
the 3 histograms of the ESS galaxies with $R_\mathrm{c}\le20.5$
classified as E/S0 (211 galaxies), Sa/Sb (88 galaxies), Sc/Ir (299
galaxies) by the cross-correlation technique, as a function of the
spectral type $\delta$. The 2 vertical lines are the corresponding
$\delta$ boundaries between the 3 spectral classes corresponding to
morphological types E/S0, Sa/Sb, Sc/Ir, and estimated by projection of
the Kennicutt spectra onto the ESS spectral sequence \citep[see \fg 2b
of][]{lapparent03a}. Figure \ref{cross_delta} shows that the ESS E/S0
cross-correlation class (solid line histogram) widely overlaps with
galaxies of Sa/Sb spectral type, which acts as a contamination of the
E/S0 LF measured from cross-correlation types; moreover, the Sa/Sb
cross-correlation class (dotted line histogram) describes a $\delta$
interval which is nearly fully included into that described by the
E/S0 histogram. This effect could explain how the Autofib survey fails
to measure the early-type intrinsic LF, and why its red-E, blue-E, and
Sab LFs have similar faint-end slopes. In a similar fashion, \fg
\ref{cross_delta} shows that spectral types Sa/Sb significantly
contaminate the cross-correlation types Sc/Ir: this could provide an
interpretation of the similar shape parameters $M^*$ and $\alpha$
measured for the Sbc, Scd and Sdm-Starburst LFs in the Autofib survey.
This analysis shows that a more robust spectral classification is
obtained by a PCA classification as used for the ESS, or by
least-square fit of the SEDs to spectral libraries, as used in the
CNOC2 survey, rather than by cross-correlation with templates.

Note that the suspected presence of type mixing in the Autofib
spectral classes complicates the interpretation of the evolution in
these LFs, parameterized in Table 2 of \citet{heyl97}.  When
calculating the Autofib Schechter parameters for $z=0.5$ (using Table
2 in \citealt{heyl97}), I derive a wider interval of faint-end slope
$-1.75\le\alpha\le-0.40$, with the following individual values:
$-0.40$ for Red-E, $-0.45$ for Blue-E, $-1.99$ for Sab, $-1.28$ for
Sbc, $-1.54$ for Scd, $-1.75$ for Sdm-Starburst. Except for the Sab
galaxies, which appear to have an anomalously large evolution rate in
$\alpha$, the values of $\alpha$ for the other classes are remarkably
close to those measured from the 5 spectral-class 2dFGRS1 LFs; the
range of $M^*$ values described by the Autofib at $z=0.5$ are also
comparable to those for the 2dFGRS1, except for the Autofib Sab
galaxies. Because the 2dFGRS1 has $z_\mathrm{max}\simeq0.15$, this
comparison casts some doubts onto the detected evolution in the
Autofib intrinsic LFs \citep{heyl97}.

In contrast with the Autofib survey and those with $0.15\la
z_\mathrm{max}\la 0.20$ (lower-right panel of \fg
\ref{mstar_alpha_B_tab}), for which $\alpha$ for all galaxy types is
steeper than $\sim-0.5$, the SAPM, DARS1, and DARS2 surveys taken
together describe the same range in $\alpha$ as the deep surveys (with
$z_\mathrm{max}\sim 0.6$; lower-left panel of \fg
\ref{mstar_alpha_B_tab}), with some LFs having values of $\alpha$ in
the interval $-0.5\la\alpha\la0.5$.  For the 3 surveys, $\alpha$
steepens for later types.  However, the value of $\alpha$ varies by at
least $0.5$ from survey to survey for a given class (see Table
\ref{mstar_alpha_B_tab}). As for the CNOC1 and CFRS $B$ LFs, based on
2 galaxy classes, $M^*$ is brighter for later types in both the DARS1
and DARS2.  Examination of the corresponding curves
\citep{efstathiou88a,metcalfe98} confirms that again, this is due in
part to the correlation between $M^*$ and $\alpha$: the bright-end of
the late-type LFs is fainter than for the early-type LF by $\sim
0.1-0.2 ^\mathrm{mag}$ for the DARS1, and by $\sim0.5 ^\mathrm{mag}$ for
the DARS2. As for the CNOC1 and CFRS $B$ LFs, the small change in the
bright-end of the LFs, characterized by an ``inverted'' $M^*$
behavior, appears to be caused by the use of only 2 classes, which
fails in separating the blue low luminosity galaxies from the luminous
Spiral galaxies.  The late-type class in the DARS1, which contains
Spiral to Irregular (denoted Irr) galaxies, might also be deficient in
blue low luminosity galaxies, which contributes to the dimming of
$M^*$: these galaxies have a lower surface brightness, and are
difficult to detect and classify visually. The weaker brightening in
$M^*$ for the Sp-Irr galaxies in the DARS2, which marks a larger
dimming of the LF bright-end for later types, may result from the
separation of the 2 classes using rest-frame color instead of the
morphological types used in the DARS1, and from the use of aperture
magnitudes for the DARS2, in replacement of the isophotal magnitudes
in the DARS1.

Although the SAPM LFs for E-S0 and Sp-Irr morphological types \resp
\citep{loveday92} do detect a $0.25 ^\mathrm{mag}$ dimming of $M^*$
for the later class, the shift between the bright-ends of the 2 LFs is
$\sim0.2 ^\mathrm{mag}$, as small as for the DARS1 survey: here, $M^*$
does not display an ``inverted'' behavior because the Sp-Irr LF has
$\alpha=-0.8$, which implies that $M^*$ reflects the location of the
bright-end (see \sct \ref{Vcomp}). The fact that the SAPM LF for
Sp-Irr galaxies fails to detect the expected steep slope for the Irr
galaxies, may be due in part to the use of only 2 sub-samples, and
also to incompleteness: the total SAPM spectroscopic sample amounts to
1658 galaxies, among which 1310 were classified as E, S0, Spiral or
Irr from visual examination of the photographic plates; one may
suspect that the 348 unclassified galaxies contain predominantly low
surface brightness objects, as these are more difficult to classify
visually. Among the low surface brightness galaxies are the late-type
low luminosity galaxies (Sd, Sm, Irr), which are the major
contributors to the steep faint-end slope of the late-type LF.
The SAPM intrinsic LF for the galaxies with strong H$\alpha$
emission line \citep[][sub-sample with EW(H$\alpha$) $\ge 15$\AA~in
Table \ref{mstar_alpha_B_tab}]{loveday99},\footnote{A classification
based on the EW[OII] is also obtained by the authors, and yields
similar results. I however favor the results based on H$\alpha$ as
this line provides a better indicator of the current star formation
rate \citep{kennicutt92b,charlot01}.} does have a steep faint-end
slope $\alpha=-1.28\pm0.30$; this improvement may be due to the
significant sample of galaxies in this class which are not
morphologically classified (see notes of Table \ref
{mstar_alpha_B_tab}), and as mentioned above, might be preferentially
Sd, Sm and Irr galaxies, those contributing to the steep faint-end
slope.

A remarkable result is that the SAPM succeeds in detecting the sharp
fall-off at faint magnitudes of the LF for morphological types E-S0,
which is characterized by $\alpha=0.20\pm0.35$ (note that the above
mentioned incompleteness would not bias this result). From visual
examination, it appears that a Gaussian LF might actually provide a
good fit to the SAPM E-S0 LF.  This confirms the reliability of the
APM morphological classification for the E-S0 galaxies, despite some
scatter, as tested by \citet{naim95}.  I emphasize that among the
redshift surveys to intermediate redshifts ($0.02\le
z_\mathrm{max}\le0.2$, shown in the 2 upper panels and in the
lower-right panel of \fg \ref{mstar_alpha_B}), the SAPM it is the
\emph{only} survey which has such a large value of $\alpha$ for the
early-type LF, in agreement at the \onesig with the values of the $B$
early-type LFs for the CFRS ($\alpha=0.00\pm0.20$), CADIS
($\alpha=0.18\pm0.22$), CNOC2 ($\alpha=0.08\pm0.14$), and ESS
($\alpha=-0.24\pm0.33$).

In contrast, the nearly flat slopes $\alpha=-0.75\pm0.28$, and
$\alpha=-0.72\pm0.29$ of the LFs for the SAPM galaxies with low and
intermediate EW(H$\alpha$) \respn, are symptomatic of type mixing, as
in the ESP LF of low [OII]-emission galaxies. It is also noticeable
that the SAPM LFs based on the equivalent width H$\alpha$ are the only
emission-line LFs which show simultaneously 2 properties of the local
intrinsic LFs: (i) a significant dimming in $M^*$ between the
early-type and late-type galaxies (namely $\simeq0.5 ^\mathrm{mag}$);
(ii) a steep faint-end slope for the late-type galaxies; these 2
properties are \emph{not} observed simultaneously in either the LCRS
$r$ LFs (see \sct \ref{Rcomp}), the ESP or the Norris $B$ LFs.  This may
be due to the joint effect of using 3 classes together with the
H$\alpha$ line, whereas the other surveys (LCRS, ESP, Norris) use only
2 classes and the [OII] line.

\subsubsection{$B$ luminosity functions at redshifts below 0.03        \label{Bcomp_low}}

Finally, the upper-left panel of \fg \ref{mstar_alpha_B} shows the
intrinsic LFs for the following nearby redshift survey (with
$z_\mathrm{max}\la 0.03$): the Nearby Optical Galaxy survey \citep[
denoted NOG]{marinoni99}; the Center for Astrophysics Redshift Survey
to $B_\mathrm{Zw}\le14.5$ \citep[CfA1]{davis82a}, complemented by the
first 2 slices of the extension to $B_\mathrm{Zw}\le15.5$ \citep[]
[the combination of the 2 surveys is denoted CfA2S]{marzke94b}; the
Southern Sky Redshift Survey \citep[denoted SSRS2]{marzke98}. For the
CfA2S and SSRS2, conversion of $M^*$ measured as Zwicky magnitude
$B_\mathrm{Zw}$ into a Johnson $B$ magnitude is based on $b_\mathrm{J}
= B_\mathrm{Zw} - 0.35$ from \citet{gaztanaga00}, on the
$B-b_\mathrm{J}=0.28(B-V)$ color equation determined by \citet[][see
also \citealt{norberg02}]{blair82}, and on the $B-V$ colors calculated
by \citet[Table 3a]{fukugita95}: for the 5 CfA2S morphological types,
I use the $B-V$ colors 0.96, 0.85, 0.78, 0.50, 0.27, 0.04, for listed
types E, S0, Sab, Scd, Im \respn; for the 3 SSRS2 morphological types,
I use the average $B-V$ color $0.905$ over listed types E and S0, and
0.57, 0.27 for listed types Sbc, Im respectively.  The resulting
$B-b_\mathrm{J}$ colors are 0.27, 0.24, 0.22, 0.14, 0.08 for listed
types E, S0, Sab, Scd, Im \respn, assigned to the CfA2S; 0.25 for the
average between listed types E and S0, and 0.16, 0.08 for listed types
Sbc, Im \respn, assigned to the SSRS2.

The NOG, CfA2S and SSRS2 redshift surveys are all based on galaxy
catalogues extracted from photographic plates
\citep{zwicky,nilson73art,eso,gsc}, and the intrinsic LFs are based on the
morphological types in the revised Hubble classification scheme
\citep[see][]{rc3}. The morphological classes corresponding to the
plotted points are indicated in Table \ref{mstar_alpha_B_tab}.  The
variations in [$M^*$,$\alpha$] for the 3 surveys resembles those for
the SDSS-Morph survey in the $R_\mathrm{c}$ band: only the LFs for the
latest class, corresponding to Sm-Im galaxies in the CfA2S and NOG
surveys, and to Irr-Pec galaxies in the SSRS2, show a clear steepening
of $\alpha$, whereas the LFs for earlier types have
$-1.0\la\alpha\la-0.5$. The steep faint-end slopes measured for the
Sm-Im LFs in the NOG ($\alpha=-2.41\pm0.28$) and the CfA2S
($\alpha=-1.87\pm0.15$), and the Irr-Pec LF in the SSRS2
($\alpha=-1.81\pm0.24$) suggest that the field LF for Sm-Im galaxies
might be on the average as steep or stepper than in the Centaurus
cluster, for which $\alpha=-1.35$ (see Table \ref{LFlocal_tab} in \sct
\ref{LFlocal}). However, in the 3 samples, the latest class does not
show the dimming in $M^*$ detected in the ESS, CADIS and COMBO-17 $B$
LFs (see lower-left panel of \fg \ref{mstar_alpha_B}), and in the
SDSS-Morph LFs converted into the $R_\mathrm{c}$ filter (see left
panel of \fg \ref{mstar_alpha_R}), which is caused by a dominating
population of dI galaxies in these classes: there is no change in
$M^*(B)$ from the Sc-Sd to the Sm-Im LF in the CfA2S, a $0.73
^\mathrm{mag}$ brightening from the Sc-Sd to the Sm-Im LF in the NOG,
and a $0.39 ^\mathrm{mag}$ brightening from the Spiral to the Irr-Pec
LF in the SSRS2.  The magnitude difference between the peak magnitude
$M_0$ of the Sc Gaussian LF and the Schechter $M^*$ for the dI
galaxies is $1.7 ^\mathrm{mag}$, $1.5 ^\mathrm{mag}$ and $1.4
^\mathrm{mag}$ in the $R_\mathrm{c}$, $V$ and $B$ bands \resp (see
Table \ref{LFlocal_tab}), and is therefore expected to be detectable
in all 3 filters. The absence of a shift towards fainter values of
$M^*$ between the Sc-Sd/Spiral LFs and the Sm-Im/Irr-Pec LFs in the
NOG, CfA2S and SSRS2 surveys suggests that the Sm-Im/Irr-Pec classes
in these surveys might be contaminated by galaxies of earlier
morphological type, and thus higher luminosity.

Although the NOG, CfA2S and SSRS2 probe the LFs to $M(B)\simeq-14.5$,
nearly $2 ^\mathrm{mag}$ fainter than in the SDSS-Morph, neither the
CfA2S nor the SSRS2 detect the Gaussian shape of the intrinsic LFs
measured for local E and S0 galaxies, which would be characterized by
a fall-off the LFs by a factor 10 or more at $M(R_\mathrm{c})\ga-17$
(see \fg \ref{LFlocal_fig}), that is $M(B)\ga-16$ using the
$B-R_\mathrm{c}$ color of an Sbc galaxy \citep{fukugita95}: both the
CfA2S and the SSRS2 measure nearly flat slopes for the E and S0 LFs
out to $M(B)\simeq-14.5$, similarly to the SDSS-Morph LFs in the
$R_\mathrm{c}$ band. The CfA2S and SSRS2 E-S0 sub-samples might
therefore contain a contribution from dSph galaxies, in a similar
fashion as the E-S0 LF in the SDSS-Morph sample \citep[][see \sct
\ref{Rcomp}]{nakamura03}. Only the NOG obtains a bounded behavior at
faint magnitudes for the E galaxy LF, with $\alpha=-0.47\pm0.22$. A
faint-end bounded LF is also measured for the NOG Sa-Sb galaxies.

At last, the LFs for giant Spiral galaxies (Sa, Sb, Sc) in the CfA2S
and SSRS2, and for the S0 and Sc-Sd galaxies in the NOG all have
nearly flat slopes. These sub-samples might also contain a
contribution from dwarf galaxies, in a similar fashion as the
ESO-Sculptor intermediate-type LF \citep{lapparent03a}.  The NOG,
CfA2S and SSRS2 morphological classifications are also likely to be
subject to some amount of type mixing, due to the dispersion in visual
classification techniques \citep{lahav95}.

\subsection{$I$ band          \label{Icomp}}

I found no measurement of intrinsic LFs in the $I$ band. Despite the
$I$ selection of the CFRS \citep{lilly95} and CADIS samples
\citep{fried01}, only $B$ intrinsic LFs are measured for these
samples. 

\subsection{SDSS and Durham general luminosity functions  \label{genLF}}

For comparison of the intrinsic LFs with the unique measurement of the
``general'' LF in a given band, I plot in \fgs \ref{mstar_alpha_UV},
\ref{mstar_alpha_R} and \ref{mstar_alpha_B} the Schechter parameters
of the general LFs for: the Sloan Digital Sky Survey \citep[denoted
SDSS]{blanton03}, which provides $u^*g^*r^*i^*z^*$ measurements in the
SDSS filters \citep{fukugita96} blueshifted by $z=0.1$; and the
Durham-Anglo-Australian-Telescope Redshift Survey
$UBVR_\mathrm{c}I_\mathrm{c}$ re-measurements \citep[denoted
DARS2]{metcalfe98}. The SDSS and DARS2 surveys provide the only 
5-band multi-color measurements of the general LF in the optical. They
may therefore serve as reference for comparison among the different
filters.  Because the DARS2 LF is split into 2 color-based intrinsic
LFs in the $B$ band, I do not plot its general LF in this filter.  I
show instead the $B$ general LF measured from the
Durham-UK-Schmidt-Telescope Redshift Survey \citep[denoted
DUKST]{ratcliffe98a}.

The SDSS general LFs in the $u^*g^*r^*i^*$ bands \citep{fukugita96}
are converted into the $UBVR_\mathrm{c}I_\mathrm{c}$ bands using the
following transformations, based on the colors of an Sbc galaxy
\citep[][Tables 3a and 3m; here we assume that the color changes
occurring when blueshifting the SDSS filters by $z=0.1$ are
negligeable]{fukugita95}:
\begin{equation}
\label{sdss}
\begin{array}{ll}
M^*(U)            = M^*(u^*) - 0.82&;\, \alpha(U)            = \alpha(u^*) ;\\
M^*(B)            = M^*(g^*) + 0.34&;\, \alpha(B)            = \frac{\alpha(u^*)+\alpha(g^*)}{2} ;\\
M^*(V)            = M^*(g^*) - 0.23&;\, \alpha(V)            = \alpha(g^*) ;\\
M^*(R_\mathrm{c}) = M^*(r^*) - 0.23&;\, \alpha(R_\mathrm{c}) = \alpha(r^*) ;\\
M^*(I_\mathrm{c}) = M^*(i^*) - 0.51&;\, \alpha(I_\mathrm{c}) = \alpha(i^*) ;\\
\end{array}
\end{equation}
Note that the uncertainty in $\alpha(B)$ is estimated as
$\sqrt{\sigma_\alpha(u^*)^2+\sigma_\alpha(g^*)^2}/2$, as implied by
\eq \ref{sdss} above. The $M^*$ value in the $b_\mathrm{J}$ band
measured for the DUKST is converted into the $B$ band using the
intermediate $B-V=0.57$ intermediate color for listed type Sbc in
\citet{fukugita95}, and the $B-b_\mathrm{J}=0.28(B-V)$ color equation
determined by \citet[][see also \citealt{norberg02}]{blair82},
yielding $B-b_\mathrm{J}=0.16$.  The resulting Schechter parameters
for the SDSS, DARS2, and DUKST are listed in Tables
\ref{mstar_alpha_UV_tab} to \ref{mstar_alpha_B_tab} (the parameters
for the $I$ LFs are listed at the end of Table
\ref{mstar_alpha_R_tab}).

Figures \ref{mstar_alpha_UV}, \ref{mstar_alpha_R} and
\ref{mstar_alpha_B} show that when compared with the Schechter
parameters for the intrinsic LFs in the same filter, the general LFs
for the DARS2 (and the DUKST) have values of $M^*$ comparable or
brighter than the values among the intrinsic LFs. This is in agreement
with the expectation that in a general LF, $M^*$ is principally
determined by the most luminous galaxies in the sample.  In contrast,
the SDSS LFs have values of $M^*$ $\sim1^\mathrm{mag}$ fainter than
in the DARS2 (in $U$, $V$, $R_\mathrm{c}$) thus lying at the median or
faintest values of $M^*$ measured for the intrinsic LFs at similar
redshifts. This difference may be due to the fact that redshift
evolution is accounted for in the derivation of the SDSS LFs, whereas
this is not the case in the other surveys with $z_\mathrm{max}\la
0.2$. Moreover, the general LFs for the SDSS, DARS2, and DUKST have
flat or slightly steeper slopes ($-1.04 \le \alpha \le -0.90$ for the
SDSS and DUKST; for the DARS2, $\alpha=-1.20$ is fixed to the value
measured in the $B$ band), whatever the range of $\alpha$ measured for
the intrinsic LFs of the various surveys in the corresponding
filter. I have shown above how type mixing inevitably results in a
nearly flat faint-end slope. This also applies to the general LF.

The general LF over a complete region of the Universe, as sampled in
systematic redshift surveys, thus provides no direct indication on the
Gaussian nature of the LFs for the giant galaxies, and on a steep
faint-end slope for the dwarf galaxies.  Given the variety of
intrinsic LFs described in \sct \ref{LFlocal}, it is a priori
surprising that the mixing of all galaxy types in redshift surveys
results in a general LF which is well fit by a Schechter function. In
local surveys of galaxy concentrations, the contributions from giant
and dwarf populations are both detected, and their signatures are an
exponential fall-off at bright magnitudes, and a steep power-law
behavior a faint magnitudes \respn, with a plateau or a knee in the
intermediate regime \citep[see for example][]{trentham02b}. Similar
behaviors are detected in general LFs for clusters of galaxies at
higher redshift \citep{driver94,wilson97,trentham98,garilli99,
durret00,beijersbergen02a,yagi02b,mobasher03}.  In systematic
redshift surveys over a given region of the Universe, the contribution
from galaxy concentrations is still present. It is however
complemented by the contribution from the numerous field Spiral
galaxies, existing in a larger proportion than in groups and
clusters. It it likely that the field Spiral galaxies cause a
significant increase in the general LF at intermediate magnitudes,
thus ``filling-in'' the mentioned ``plateau'' region, and making the
Schechter form an adequate description over more than 5 magnitudes.
Then, as suggested by \citet{binggeli88} and \citet{ferguson91}, the
variations of $M^*$ and $\alpha$ in the general LF as a function of
sample and filter may simply reflect the average proportions of the
various galaxy types in the survey region.

\section{Conclusions and prospects     \label{concl}}

I perform a detailed comparison of all the existing measurements of
intrinsic LFs in the optical domain, derived from redshift surveys
with effective depth $z\simeq0.03$ to $0.6$ and converted into the
$UBVR_\mathrm{c}I_\mathrm{c}$ system wherever necessary. The shape of
the various LFs is compared among the different surveys and galaxy
classes, using the Schechter parameters $M^*$ and $\alpha$.  In this
comparison, I use as reference the intrinsic LFs per morphological
type measured from local galaxy concentrations
\citep{sandage85b,jerjen97b}.

Each survey detects variations in the shape of the LF with galaxy
type. However, the LFs for a given galaxy type widely vary from survey
to survey. I interpret these differences in terms of the
classification schemes for defining the galaxy classes (based on
morphological types, spectral types, cross-correlation types, colors,
or equivalent width of emission lines), and show that they often
induce some mixing of distinct morphological types, which in turn
complicates the interpretation of the LFs.

The salient results which I emphasize or derive in the present
analysis are:
\begin{itemize}

\item spectral classification with accurate flux calibration and a
minimum of 3 classes allows to observe both the Gaussian early-type LF
(corresponding to E, S0, and sometimes Sa galaxies), and the dimming
of the late-type LF (containing usually Sc, Sd/Sm and Irr galaxies) as
illustrated by the CNOC2 \citep{lin99} and ESO-Sculptor Survey
\citep{lapparent03a}.

\item the nearby Center for Astrophysics Redshift Survey
\citep{marzke94b}, the Southern Sky Redshift Survey \citep{marzke98},
and the deeper sample extracted from the SDSS Early Data Release
\citep{nakamura03}, all based on visual morphological classification,
detect a nearly flat faint-end slope for their earliest-type LFs, thus
failing to detect the Gaussian E/S0 LF; moreover, these surveys fail to
detect the dimming of the Sm-Im LF compared to the Sc-Sd LF.

\item the Autofib \citep{heyl97} and the 2 preliminary samples of the 2dF
Galaxy Redshift Survey samples \citep[][2dFGRS]{folkes99,madgwick02a},
all based on spectral classification, also fail to detect the
Gaussian LF for E/S0.

\item although the Autofib survey \citep{heyl97} is based on slit
spectroscopy, a fair amount of type mixing appears to bias the derived
LFs, because of inaccuracies in the spectral classification which is
based on cross-correlation with galaxy templates; the effect is
demonstrated using the ESO-Sculptor Survey \citep{lapparent03a}, for
which both cross-correlation types and PCA (Principal Component
Analysis) spectral types are available.

\item the continuous variation in the Schechter faint-end slope
of the spectral-type LFs measured in $b_\mathrm{J}$ for the 2 preliminary
samples of the 2dFGRS \citep{folkes99,madgwick02a}, and in the 6
spectral-type LFs measured in $R_\mathrm{c}$ for the Las Campanas
Redshift Survey \citep{bromley98} is interpreted as type mixing
between the giant galaxies with Gaussian LFs and the dwarf galaxies
with Schechter LFs: this partly results from the aperture and
flux-calibration biases affecting redshift surveys obtained with
multi-fiber spectrographs.

\item when LFs are measured for 2 sub-samples separated by color (as in the
Canada-France Redshift Survey, \citealp{lilly95}; the CNOC1 survey,
\citealp{lin97}; and the Century Survey, \citealp{brown01}), or
separated by the equivalent width of characteristic emission lines (as
in the ESO Slice Project, \citealt{zucca97}; the Norris survey,
\citealp{small97b}; and the Stromlo-APM survey, \citealp{loveday99};
see also \citealt{lin96}) they are insufficient for estimation of the
intrinsic LFs as they not only fail to separate the various
populations of giant and dwarf galaxies, but they also mix giant
galaxies of different morphological type; this effect is illustrated
using the ESO-Sculptor Survey \citep{lapparent03a}, for which PCA
spectral types, colors and equivalent width of [OII] emission are
available.

\item although the COMBO-17 LFs in the $B$ band \citep{wolf03} are
consistent with those from the comparable CADIS \citep{fried01}, and
with those from the CNOC2 \citep{lin99} and ESO-Sculptor
\citep{lapparent03a} surveys, the COMBO-17 LFs converted into the $U$
and $R_\mathrm{c}$ bands shows significant differences with the CNOC2
and ESO-Sculptor for the intermediate spectral types corresponding to
Spiral galaxies.  This may result from the complex selection effects
inherent to the use of medium-band photometry for redshift measurement
in the COMBO-17 survey, and/or from its color transformations
from the $r^*$ and $m_{280}$ bands into the $R_\mathrm{c}$ and $U$
bands \resp \citep{wolfp}.

\end{itemize}

One conclusion which I draw from these various results is that the
spectral classifications used in the CNOC2 \citep{lin99} and
ESO-Sculptor \citep{lapparent03a} surveys, both based on multi-slit
spectroscopy, provide the least biases estimates of intrinsic LFs. The
CADIS \citep{fried01} and COMBO-17 \citep{wolf03} surveys, based on
photometric redshifts using medium-band filters, also provide
consistent intrinsic LFs in the $B$ band with the CNOC2 and
ESO-Sculptor.  The 4 mentioned surveys are based on CCD photometry
combined with a spectral classification with accurate flux
calibration, which therefore appears, among the mentioned surveys, as
the optimal combination for estimating the intrinsic LFs.  The
systematic effects affecting a spectral classification based on
multi-fiber spectroscopy as in the 2dF Galaxy Redshift Survey
\citep{folkes99,madgwick02a} cause type mixing among the various
morphological classes which significantly biases the estimates of
intrinsic LFs.

Surprisingly, spectral classification at $z\sim0.5$ provides better
estimates of the intrinsic LFs than the first generation of redshift
surveys to $z\la0.03$ (the Nearby Optical Galaxy survey,
\citealp{marinoni99}; the Center for Astrophysics Redshift Survey,
\citealp{marzke94b}; and the Southern Sky Redshift Survey
\citealp{marzke98}), although the latter surveys are based on direct
morphological classification. The intrinsic LFs derived from the
nearby surveys are likely to be biased in their magnitudes and
morphological classification because of (i) the non-linear response of
photographic plates, (ii) their narrow dynamic range, and (iii) the
human estimation of the magnitudes and galaxy types
\citep[see][]{lahav95}.  Visual morphological classification is indeed
largely subjective, even from good quality imaging
\citep{lahav95,abraham96b}. The similar LFs derived from the SDSS Early
Data Release \citep{nakamura03} suggest that despite the improvement
brought by CCD imaging, the visual morphological classification
performed for this sample also suffers similar biases as in the nearby
redshift surveys. Only the $B$ LF for the Stromlo-APM survey
\citep[denoted SAPM]{loveday92} detects the Gaussian behavior for the
E-S0 galaxies. This may however be due to the difficulty of
classifying low surface brightness galaxies, and the possible resulting
incompleteness of the SAPM classification in faint early-type galaxies.

Another noticeable result is that no existing redshift survey with
morphological, spectral or color classification has measured the
bounded LFs for the individual Spiral types (Sa, Sb, Sc, Sd), nor the
Gaussian shape of the LF for Spiral galaxies altogether, as measured
locally \citep{sandage85b,jerjen97b}.  This confirms the
interpretation of \citet{lapparent03a}, who show that the ESO-Sculptor
Survey spectral-type LFs corresponding to Spiral galaxies might
contain at their faint end a contribution from early-type dwarf
galaxies.  Failure to separate the giant and dwarf galaxy populations
in all existing redshift surveys, whatever the classification
criterion, thus prevents any reliable measure of the Spiral intrinsic
LFs. Even the 2 preliminary samples of the 2dFGRS
\citep{folkes99,madgwick02a} and the commissioning data from the Sloan
Digital Sky Survey \citep{blanton01}, which both sample with high
statistical significance the galaxy distribution to $M(B)\simeq -16$,
fail to measure the intrinsic Gaussian LF for Spiral galaxies.

The present analysis therefore emphasizes the need for more reliable
and systematic approaches for morphological classification from
$z\simeq0$ to $z\ga1$, indispensable for measuring the intrinsic LFs,
and their possible evolution with redshift.  Morphological
classification at $z\ga0.5$ is however a delicate task: it is
wavelength dependent \citep{burgarella01,kuchinski01}, and detected
evolution in galaxy morphology at $z\ga1$ complicates the definition
of reference types \citep{vandenbergh97,vandenbergh00,vandenbergh01}.
A reliable discrimination among the morphological types is crucial for
estimating intrinsic LFs, as these show a wide variety of shape and
characteristic parameters for the different galaxy types (see \sct
\ref{LFlocal} and Table \ref{LFlocal_tab}).  Inaccurate classification
may then cause biases in the derived intrinsic LFs.  In the analysis
of the ESO-Sculptor Survey LFs, \citet{lapparent03a} suggest that a
useful morphological classification for measuring intrinsic LFs could
include the surface brightness profile of the galaxies, as it allows
to separate giant and dwarf galaxies, which have markedly different
intrinsic LFs.  So far, none of the existing redshift surveys provide
separate LF measurements for the giant and dwarf galaxies.

Derivation of the intrinsic LFs per morphological type will also
require redshift samples with at least $\sim 10^5$ galaxies, in order
to have sufficient statistical samples of the various giant and dwarf
galaxy types.  Such large sample shall be obtained at $z\la0.2$ by the
Sloan Digital Sky Survey (see http://www.sdss.org/), and the 2dF
Galaxy Redshift Survey (see http://www.mso.anu.edu.au/2dFGRS/). The
preliminary LF measurements from these surveys which are analyzed here
show that both the spectral classification used in the 2dF survey
(based on a PCA spectral classification, \citealp{madgwick02a}, and
interpreted in terms of star formation history,
\citealp{madgwick02b}), and the visual morphological classification
used in the Early Data Release of the SDSS \citep{nakamura03} appear
insufficient for measurement of the intrinsic LFs.  In contrast, the
LFs derived from the SDSS commissioning data and based on 5 intervals
of $g^*-r^*$ color succeed in detecting the Gaussian LF for the giant
early-type galaxies \citep{blanton01}.  Note that in view of the
analyses presented here, the 2 color classes separated by
$u^*-r*=2.22$ and shown by \citet{strateva01} to split SDSS galaxies
according to morphological type and radial profile are likely to be
insufficient to recover either the Gaussian LFs for the giant galaxies
or the Schechter LFs for the dwarf galaxies.  As I have shown here,
measurement of LFs based on 2 color sub-samples lacks the necessary
discriminatory power necessary for detecting the variations in
luminosity as a function of morphological type which are traced by the
intrinsic LFs.

I therefore recommend that in the case of multi-fiber surveys to
moderate depths ($z_\mathrm{max}\la 0.2$), galaxy classification for
estimation of the intrinsic LFs be based on rest-frame colors
\emph{rather} than on the spectral data.  Whereas the fiber spectra
only sample partial regions of the objects, the rest-frame colors do
include the full light from the objects, and the accuracy of the
photometric calibrations ensures that the colors reflect the shape of
the SED for each object. A combined approach, which might yield
improved results over a classification based on either multi-fiber
spectroscopy or rest-frame color, is the calibration of spectral data
using multi-color photometry, and its subsequent spectral
classification.  In the SDSS, the accurate photometry based on CCD
multi-color imaging over 5 optical bands \citep{fukugita96} should
allow one to obtain such an improved classification. Performing such an
analysis for 2dF Galaxy Redshift Survey, which is based on the APM
scans of UK Schmidt $J$ photographic plates
\citep{maddox90a,maddox90b}, with no color information, would require
additional photometry in another band. Neither rest-frame colors nor
spectral classification are however sufficient to separate the giant
and dwarf galaxy populations and to measure the intrinsic LFs. To meet
this goal, one additional step is the availability of the surface
brightness profiles for all galaxies. A recent analysis of galaxy
properties in a SDSS sample does use the radial profile of the objects
\citep{blanton02}.

The DEEP2 \citep{davis02} and VIRMOS \citep{lefevre01} surveys, using
deep CCD imaging and efficient multi-slit spectrographs on the Keck
telescopes \citep{cowley97,james98} and ESO-VLT \citep{lefevre01}
\respn, are also expected to bring useful measurements of intrinsic
LFs at $z\sim1$ and their possible evolution with redshift, provided
that these surveys succeed in separating the various morphological
types, including the giant and dwarf populations. So far, detection of
evolution in the intrinsic LFs are based on either too few classes to
allow a definite interpretation in terms of one morphological class
\citep[][using 2 and 3 classes]{lilly95,lin99,fried01,lapparent03a},
or on more classes but are suspected to suffer from type contamination
\citep[][see \sct \ref{Bcomp_mid_low} above]{heyl97}.  Reliable
analyses of the evolution in the intrinsic LFs with redshift will
require the best resolution in morphological types.

I emphasize that an efficient technique for measuring LFs is the use
of photometric redshifts: similar spectral type LFs are derived in the
$B$ band for on one hand the CADIS \citep{fried01} and COMBO-17
\citep{wolf03} surveys, and on the other hand the CNOC2 \citep{lin99}
and ESO-Sculptor \citep{lapparent03a} surveys despite the
significantly larger uncertainties in the redshifts derived from the
CADIS and COMBO-17, measured from a combination of wide and
medium-band filters ($\sigma(z)\le0.03$).  This demonstrates the
interest of the ``photometric redshift'' approach for measuring LFs,
which has the advantage of providing large samples at a reduced cost
in telescope time. Along this line, 2 forthcoming surveys are expected
to provide significant contributions to the measurement of the
intrinsic LFs:
\begin{itemize}
\item the Large-Zenith-Telescope (LZT) project, which aims at
  obtaining redshifts for $\sim 10^6$ galaxies to $z\la 1$ at
  $R\la23$ over 40 deg$^2$ of the sky, using 40 medium-band filters
  \citep[][see also
  http://www.astro.ubc.ca/LMT/lzt.html]{cabanac02,hickson98b};
\item the Canada-France-Hawaii Telescope Legacy Survey (see
  http://www.cfht.hawaii.edu/Science/CFHTLS/), which will cover
  $170$ deg$^2$ in the wide-band filters $u^*g'r'i'z'$ at
  $r'\la25.7$.
\end{itemize}
On one hand, the LZT will provide a detailed PCA-spectral
classification and redshift measurements with an accuracy
$\sigma(z)\le0.05$ at $z\la1$ \citep{cabanac02}, complemented by
measures of surface brightness. On the other hand, by application of
quantitative algorithms to the multi-color images obtained in
excellent imaging conditions, the CFHT Legacy Survey will allow a
detailed morphological classification, which shall be complemented by
photometric redshifts with uncertainties $\sigma(z)/z\sim0.1$
\citep{bolzonella00}.  Both surveys will provide 2 orders of magnitude
larger samples than the redshifts surveys, thus allowing useful
measurements of the intrinsic LFs to $z\la1$.

I show here that the wide majority of intrinsic LF measurements were
performed in the $B$ band, with few measurements in the $R_\mathrm{c}$
band, 2 measures in the $U$ band, a single one in the $V$ band, and no
measurement in the $I_\mathrm{c}$ band. It will be important that the
mentioned surveys under completion allow measurement of the intrinsic
LFs in a variety of filters, including the ultraviolet and
infrared. This will be a crucial step towards understanding the
contribution from the different galaxy populations to the luminosity
density of the Universe and for constraining its evolution with
redshift. 

\begin{acknowledgements}
I am grateful to St\'ephane Arnouts for encouraging me to publish this
analysis as a separate article, for suggesting its title, and for his
useful comments on the article content. I also thank Josef Fried for
providing me with the numerical values for the Schechter parameters of
the CADIS luminosity functions (used in Table \ref{mstar_alpha_B_tab}
and in \fg \ref{mstar_alpha_B}), and Chris Wolf for providing the
color corrections from the COMBO-17 filters into the Johnson system
(used in Tables \ref{mstar_alpha_UV_tab} and \ref{mstar_alpha_R_tab},
and in \fgs \ref{mstar_alpha_UV} and \ref{mstar_alpha_R}).
\end{acknowledgements}

\bibliography{lapparent2}
\bibliographystyle{aa}

\end{document}